\documentclass[iop]{emulateapj}
\usepackage{hyperref}

\slugcomment{Accepted to AJ}

\shorttitle{Robo-AO Kepler Planetary Candidate Survey II}
\shortauthors{Baranec et al.}

\begin{document}


\title{Robo-AO Kepler Planetary Candidate Survey II: Adaptive Optics Imaging of 969 Kepler Exoplanet Candidate Host Stars}


\author{Christoph Baranec\altaffilmark{1}, Carl Ziegler\altaffilmark{2}, Nicholas M. Law\altaffilmark{2},  Tim Morton\altaffilmark{4}, Reed Riddle\altaffilmark{3}, Dani Atkinson\altaffilmark{1}, Jessica Schonhut\altaffilmark{1, 5} and Justin Crepp\altaffilmark{6}}
\email{baranec@hawaii.edu}


\altaffiltext{1}{Institute for Astronomy, University of Hawai`i at M\={a}noa, Hilo, HI 96720-2700, USA}
\altaffiltext{2}{Department of Physics and Astronomy, University of North Carolina at Chapel Hill, Chapel Hill, NC 27599-3255, USA}
\altaffiltext{3}{Department of Astrophysical Sciences, Princeton University, Princeton, NJ 08544, USA}
\altaffiltext{4}{Division of Physics, Mathematics, and Astronomy, California Institute of Technology, Pasadena, CA 91125, USA}
\altaffiltext{5}{University of Hertfordshire, Hatfield, Hertfordshire AL10 9AB, United Kingdom}
\altaffiltext{6}{Department of Physics, University of Notre Dame, Notre Dame, IN 46556, USA}


\begin{abstract}

We initiated the Robo-AO \textit{Kepler} Planetary Candidate Survey in 2012 to observe each \textit{Kepler} exoplanet candidate host star with high-angular-resolution visible-light laser-adaptive-optics imaging. Our goal is to find nearby stars lying in \textit{Kepler}'s photometric apertures that are responsible for the relatively high probability of false-positive exoplanet detections and that cause underestimates of the size of transit radii. Our comprehensive survey will also shed light on the effects of stellar multiplicity on exoplanet properties and will identify rare exoplanetary architectures. In this second part of our ongoing survey, we observed an additional 969 \textit{Kepler} planet candidate hosts and we report blended stellar companions up to $\Delta m \approx 6$ that contribute to \textit{Kepler}'s measured light curves. We found 203 companions within $\sim$4\arcsec of 181 of the \textit{Kepler} stars, of which 141 are new discoveries.  We measure the nearby-star probability for this sample of \textit{Kepler} planet candidate host stars to be 10.6\% $\pm$ 1.1\% at angular separations up to 2\farcs5, significantly higher than the 7.4\% $\pm$ 1.0\% probability discovered in our initial sample of 715 stars; we find the probability increases to 17.6\% $\pm$ 1.5\% out to a separation of 4\farcs0. The median position of KOIs observed in this survey are 1.1$^{\circ}$ closer to the galactic plane which may account for some of the nearby-star probability enhancement. We additionally detail 50 Keck adaptive optics images of Robo-AO observed KOIs in order to confirm 37 companions detected at a $<5\sigma$ significance level and to obtain additional infrared photometry on higher-significance detected companions.  
\end{abstract}





\def\Kepler{\textit{Kepler}}
\section{Introduction}
The primary \textit{Kepler} mission photometrically observed approximately 200,000 stars for 4 years in the search for transiting exoplanets. As of the Q1-Q17 DR24 dataset release \citep{DR24}, there are 3,324 \textit{Kepler} Objects of Interest (KOI) stars in the \textit{Kepler} input catalog that are designated as either `CONFIRMED' or `CANDIDATE' in the NASA Exoplanet Archive \citep{Akeson13}, with 4,302 repeating transit signals indicative of transiting exoplanets. 
While \textit{Kepler} is one of the best facilities to measure the periodic dips in stellar brightness caused by transiting exoplanets, it has a coarse pixel size, $\sim4\arcsec$, and a mean 95\% encircled-energy diameter of 4.3 pixels \citep{haas10} that make it unsuitable for reliably detecting multiple sources within the \textit{Kepler} photometric apertures. Follow-up high-angular-resolution imaging of KOIs is used to determine the sources contributing to the \textit{Kepler} light curves in order to rule out astrophysical false positives (e.g., \citealt{morton11}, \citealt{dressing13}, \citealt{fressin13}, \citealt{santerne13}), to accurately measure the radii of the detected planets with respect to their host star (e.g., \citealt{C15, Ciardi15}), to measure the effect of stellar multiplicity on exoplanet formation (e.g., \citealt{W14a, W14b, W15, W15b}), and to study other astrophysical phenomena (e.g., \citealt{Muirhead2013, Montet2015}).

In 2012, we initiated the Robo-AO \textit{Kepler} Planetary Candidate Survey in an effort to systematically observe each \textit{Kepler} planet candidate host star in a consistent way with adaptive optics (AO). No AO survey of this magnitude had previously been attempted and this type of survey has only recently been made possible with the commissioning of the Robo-AO robotic laser adaptive optics system that can observe more than 200 objects in a single night \citep{Baranec2014}. We expect that the results of this survey will be used in validating candidate exoplanets, correcting the estimates of transit radii, identifying rare and interesting exoplanetary system architectures and exploring the properties and trends of exoplanets in multiple star systems.  

During our first observing season in 2012, we observed 715 KOIs with Robo-AO (\citealt{Law2014}, henceforth Paper I). Of the 715 KOIs observed, we found 53 to have a fainter stellar companion within a 2\farcs5 radius, leading to a nearby-star probability of 7.4\% $\pm$ 1.0\%. We now report results from the second part of our ongoing survey, comprising a further 969 observations of KOIs with Robo-AO. We additionally report Keck adaptive optics images of 50 of the Robo-AO observed KOIs to confirm companion detections made at low significance, $<5\sigma$, and to obtain additional infrared photometry that will later be used to better constrain photometric parallaxes to determine if the companions are physically associated (D. Atkinson et al., 2016, in preparation). Of the 969 KOIs observed, we have found 203 companions\footnote{For brevity we denote stars which we have found within our detection radius of KOIs as ``companions,'' in the sense that they are asterisms associated on the sky.} of which 141 are new discoveries.

This paper is organized as follows. In Section \ref{sec:STO} we describe the KOI survey target selection and the observations made with Robo-AO and NIRC2. Sections \ref{sec:raodata} and \ref{sec:kdata} describe the Robo-AO and NIRC2 data reduction and companion-detection pipelines respectively. In Section \ref{sec:discoveries} we describe the results of the survey, including the discovered companions, and compare these with other surveys and observations. In Section \ref{sec:conclusions} we discuss implications of the survey and how others can use these observations. Finally, we conclude with plans for current and future work in Section \ref{sec:future_work}.

\begin{figure*}
\centering
\includegraphics[width=0.83\paperwidth]{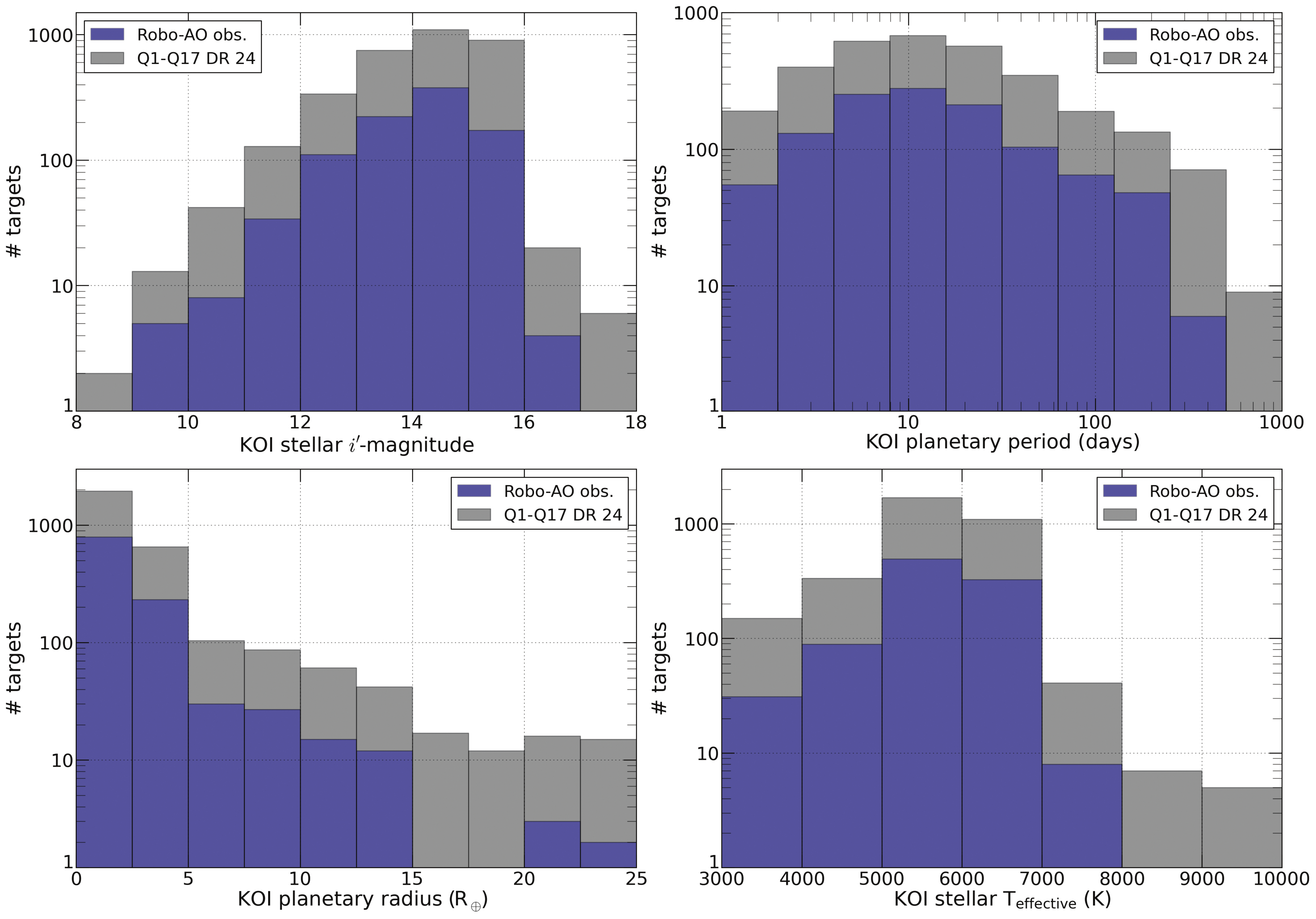}
\caption{Comparison of the distribution of the Robo-AO sample in this paper, from Q1-Q12 \citep{Rowe2015}, to the set of KOIs from Q1-Q17 DR 24 \citep{DR24}.}
\label{fig:histograms}
\end{figure*}

\section{Survey, Targets and Observations}
\label{sec:STO}

\subsection{Target Selection}

We selected targets that we had not previously observed from the KOIs Catalog based on the Q1-Q12 \textit{Kepler} data \citep{Rowe2015}. These targets were added to the Robo-AO intelligent observing queue \citep{Riddle2014} and observed during the summer of 2013. While we imposed no artificial magnitude limit on the targets, there were only 4 targets fainter than 16th magnitude that we effectively observed. In Figure 1 we compare the Robo-AO imaged KOIs to the distribution of all current KOIs \citep{DR24}. The four graphs show the comparisons in the characteristics of KOI stellar magnitude and effective temperature; and planetary orbital period and radius. The list we have observed correlates closely with the \textit{Kepler} Q1-Q17 DR24 list in all of these categories, with the exception of the longest period planets which are fewer in number in the Q1-Q12 list due to the shorter time baseline. It is only by coincidence that we did not observe the small number of KOIs with 15 - 20 R$_\oplus$ companions nor with stellar temperatures higher than 8000 K.

\subsection{Observations}
\label{sec:observations}

\subsubsection{Robo-AO}

We obtained high-angular-resolution images of 956 \textit{Kepler} planet candidate host stars over the course of 19 nights between 2013 July 21 and 2013 October 25 detailed in Table \ref{tab:obs_table} in the Appendix. We also include 13 images from 2012 that required additional confirmation of the KOI position in the Robo-AO field of view. All the observations were performed in a queue-scheduled mode in combination with other science programs using the Robo-AO autonomous laser adaptive optics system \citep{Baranec2013, Baranec2014} mounted on the robotic 1.5-m telescope at Palomar Observatory \citep{Cenko:06}. Table \ref{tab:specs} summarizes the system and survey specifications.

Each observation comprises a sequence of full-frame-transfer detector readouts of an electron multiplying CCD camera at the maximum rate of 8.6 Hz. Individual frames are later registered to correct for the dynamic image displacement of the KOI (Section \ref{sec:raodata}) that cannot be measured with the laser guide star. A total exposure integration time of 90 seconds was chosen so that close sources up to roughly 6 magnitudes fainter than the \textit{Kepler} object would be detected. For the majority of these observations a long-pass filter with a cut-on wavelength of 600 nm was used (LP600 hereafter). The LP600 filter approximately matches the \textit{Kepler} passband at the redder wavelengths whilst simultaneously suppressing the blue wavelengths. The blue wavelengths will degrade adaptive optics performance in the majority of seeing conditions. Compared to near infrared adaptive optics observations, the LP600 filter more closely approximates the direct measurement of the effects of any unresolved companions for the relevant \textit{Kepler} light curves. An $i'$-band filter \citep{York2000} was used during eight of the 2012 observations in an attempt to obtain slightly sharper images of brighter targets. A comparison of the two filters can be found in Paper I. 

There are two main factors that affect the quality of images acquired by Robo-AO: atmospheric seeing and the brightness of the target. For bright targets ($m_V<13$), in median seeing of 1\farcs1 \citep{Cenko:06}, Robo-AO can obtain images with a Strehl ratio of 9\%, and full-width at half-maximum (FWHM) of 0\farcs12 in $i'$-band. As the seeing approaches 1\farcs6, the Strehl ratio drops to 5\%. For fainter targets, i.e. $m_V>14$, there needs to be a sufficient number of photons in the diffraction limited core captured during each frame-transfer exposure for post-facto image registration techniques to maintain full acuity. Robo-AO is able to capture scientifically useful images on these fainter targets during times when the atmospheric seeing is favorable, so observations are often repeated until data of sufficient quality is obtained. We adopted the same automated routines used by Paper I to measure the actual imaging performance and to classify the targets into the imaging performance classes given in the full observations list; this classification was used with the contrast curve for each class to estimate the companion-detection performance for each target (Section \ref{sec:imageperf}).


\subsubsection{Keck adaptive optics}
We obtained images of 50 KOIs with the NIRC2 instrument behind the Keck II adaptive optics system that were previously observed with Robo-AO and had evidence of a companion. For KOIs brighter than $m_V \sim 13$ we typically used the KOI as the guide star in natural-guide-star mode, and for fainter KOIs we used the laser guide star, using the KOI as the tip-tilt-focus guide star \citep{KeckLGS1, KeckLGS2}. Observations were conducted on 2013 June 25, 2013 August 24 and 25, 2014 August 17 and 2015 July 25 in the K, Ks or Kp filters, and in the narrow mode of NIRC2 (9.952 mas pixel$^{-1}$; \citealt{Yelda10}). An initial 30 s exposure was taken for each target, and we waited for the low-bandwidth wavefront sensor to settle if the laser was used. The integration time and number of coadds per detector readout were adjusted to keep the peak of the stellar PSF counts less than 8,000 ADU per single integration (roughly half the dynamic range where sensitivity of the detector is linear), while maintaining a total exposure time of 30 s. Dithered images were then acquired with the primary centered in the 3 lowest noise quadrants using the `bxy3 2.5' command, for a total exposure time of 90 s.

\textbf{\begin{table}
\renewcommand{\arraystretch}{1.3}
\begin{longtable}{lc}
\caption{Specifications of the Robo-AO KOI survey}
\\
\hline
KOI targets    	& 969 \\
Exposure time & 90 seconds \\
Observation wavelengths & 600--950 nm \\
FWHM resolution   	& 0 $\farcs$12--0 $\farcs$15 \\
Field of view & 44\arcsec $\times$ 44\arcsec\\
Pixel scale & 43.1 mas / pix\\
Detector format       	& 1024$^2$ pixels\\
Observation dates &  2012 July 16 -- 2012 September 13 \\
 & 2013 July 21 -- 2013 October 25 \\
Targets observed / hour & 20\\
\hline
\label{tab:specs}
\end{longtable}
\end{table}}

\begin{figure*}
\centering
\includegraphics[width=0.75\paperwidth]{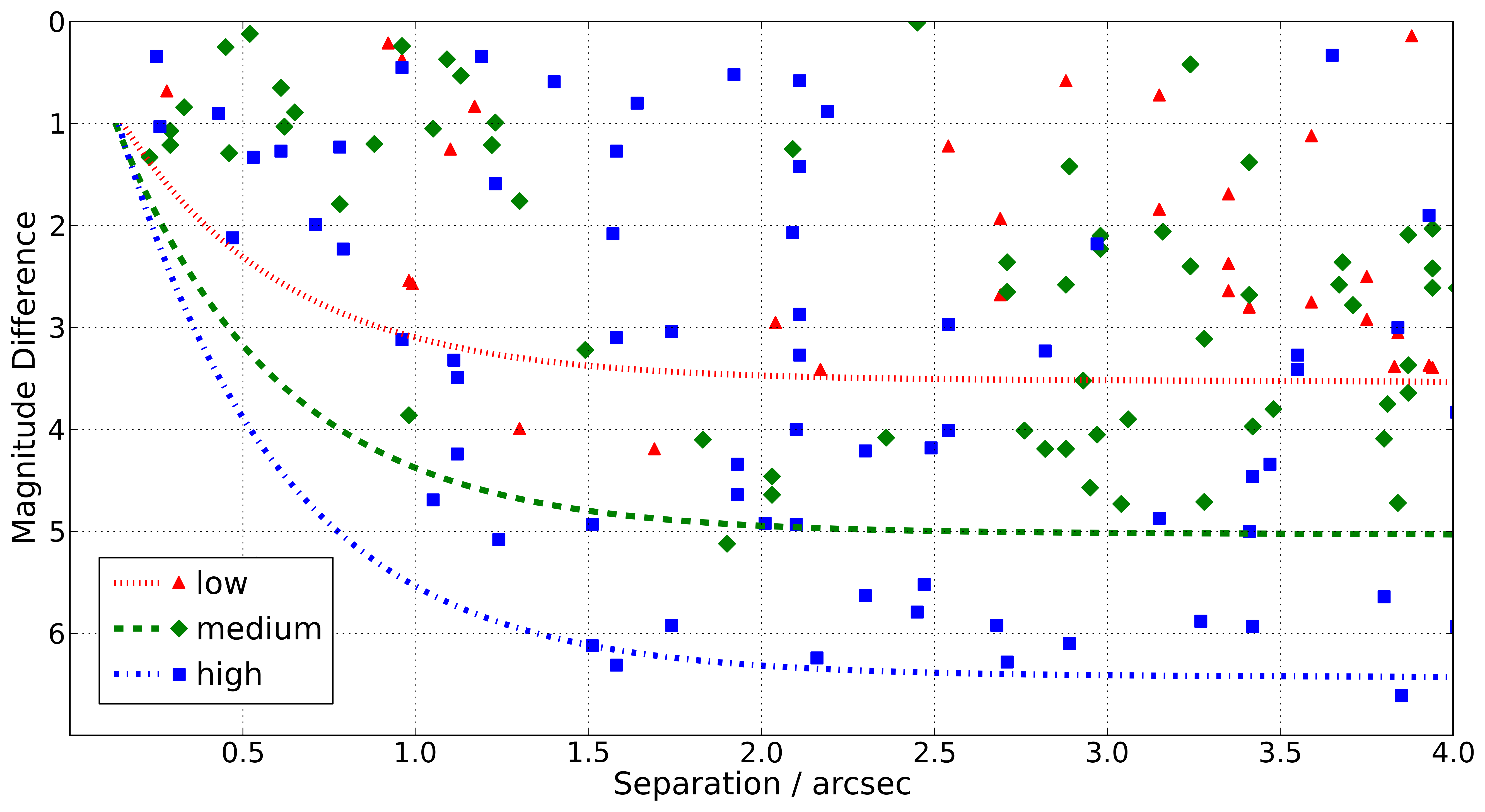}
\caption{The points on this plot show the angular separations and magnitude differences of the detected companions described in Tables \ref{tab:comp_high} and \ref{tab:comp_low}, with the color and shape of each point denoting the associated typical low-, medium- and high-performance 5$\sigma$ contrast curve during the observation (as described in Section \ref{sec:imageperf}).}
\label{fig:contrastcurves}
\end{figure*}

\section{Robo-AO Data Reduction}
\label{sec:raodata}

We adopted the same automated data reduction and analysis pipeline used in Paper I and we briefly review it here, along with minor improvements. We first manually find the location of the KOI in the field using a preliminary reduction of the data (Section \ref{sec:target_confirmation}). The pipeline first takes the short-exposure data cubes recorded by the electron multiplying CCD camera and produces dark, flat-field and tip-tilt-corrected co-added output images (Section \ref{sec:imagepipe}). We then subtract a locally optimized point-spread function (PSF) estimate from the image of the Kepler target in each field (Section \ref{sec:psfsub}), and either detect companions around the target stars or place limits on their existence (Section \ref{sec:companion}). Finally, we measure the properties of the detected companions (Section \ref{sec:compchar}).

\subsection{Target Confirmation}
\label{sec:target_confirmation}
We manually checked the location of the KOI in Digital Sky Survey (DSS) images and selected the KOI itself as the guide star to correct image displacement in each observation. In this survey, for fields where the DSS image was insufficient for KOI identification, or if the proper motion of stars made the target ambiguous, we would use the publicly available recent UKIRT J-band images of the Kepler field. For a minority of targets, there was only a single star in the field of view. For these targets, we first confirmed with UKIRT images that there were no other sources within our field; then we confirmed whether the telescope pointing offsets were stable for that particular observation by noting if prior and subsequent KOI targets landed in the same area of the detector. We note that Paper I did not include all of the observed targets in 2012 because of our inability to unambiguously identify every KOI. Using this new method of target confirmation we were able to positively identify 13 KOIs observed in 2012 and now include them in this paper.

\subsection{Imaging Pipeline}
\label{sec:imagepipe}
The Robo-AO imaging pipeline \citep{Law12, Terziev13} is based on the lucky imaging reduction system described in \citet{Law06a, Law06b, Law09}. The recorded camera frames are dark-subtracted and flat-fielded, and are then corrected for image displacement using the KOI as the reference guide star. This produced more consistent and predictable imaging performance for groups of similar KOIs, even if a brighter guide star was nearby and offered potentially increased performance.  

\subsection{PSF Subtraction}
\label{sec:psfsub}

The large number of KOI target stars observed each night are all in similar parts of the sky, have similar brightness, and were observed at similar airmasses. We take advantage of this fact and use each night's KOI observations as PSF references due to the fact that it is unlikely that a companion would be in the same position for multiple targets. We use a custom locally optimized PSF subtraction routine based on the Locally Optimized Combination of Images algorithm \citep{L7}, wherein the regions around at least 20 targets are combined to create a PSF which is an optimal local combination of the reference PSFs and is then subtracted from the target star's PSF. The PSF subtraction typically leaves residuals that are consistent with photon noise only (for these relatively short exposures). For more information and an example of the target star subtraction see Paper I.

\subsection{Automated Companion Detection}
\label{sec:companion}

To more easily and robustly find companions in this large data set, we developed a new automated companion detection algorithm for Robo-AO data, described in Paper I. During the analysis of images for this survey we extended the automated search radius from 2\farcs5 to 4\farcs0 to capture a larger population of stars that contribute to the \textit{Kepler} light curves, as well as to facilitate better comparison with other high-angular-resolution imaging surveys.

We also manually checked each image for companions after the automated companion detection to assess the performance of the automated system and to search for faint but real companions that could have been removed by spurious speckles in the PSF references. Only those that had a measured significance of $>2.0\sigma$ are reported here despite the fact that others may have confirmed their existence (e.g., the $\sim$0\farcs5 companions detected near KOI-3284 and KOI-3309). While not comprehensive, we also flagged several low-significance companions out to $\sim$4\farcs5 that were just outside of the automatic search radius during the manual check.

\subsection{Imaging Performance Metrics}
\label{sec:imageperf}

In Paper I, we evaluated the contrast-versus-radius detection performance of the PSF-subtraction and automated companion detection code by performing Monte Carlo simulations of artificial companion injection and recovery. We found that if we fit two Moffat functions to each PSF, one tuned to the PSF core and the other to the uncorrected halo, that the width of the PSF core size alone was an excellent predictor of contrast performance. On this basis, we used the PSF core size to assign targets to contrast-performance groups: `low', $<$0\farcs1, N=355; `medium', $[0\farcs1, 0\farcs14)$, N=308; and `high', $\geq0\farcs14$, N=306. Figure \ref{fig:contrastcurves} shows smoothed contrast curves resulting from the Monte Carlo companion-detection simulations for the three ranges of contrast-performance groups.

\subsection{Companion Contrast Ratios, Separations and Position Angles}
\label{sec:compchar}

We determined the contrast ratio between the companions and primaries in two ways: for the widest separations we performed aperture photometry on the original images; for the closer systems we used the estimated PSF to remove the blended contributions of each of the stars before performing aperture photometry. In all cases the aperture sizes were optimized for the system separation and the available signal. We calculated the contrast ratio uncertainty on the basis of the difference between the injected and measured contrasts of the artificial companions during the contrast-curve calculations (Section \ref{sec:companion}). We found that the detection significance of the companion was the best predictor of the contrast ratio accuracy, and so we use that relation to estimate the contrast ratio uncertainty for each companion. We note that the uncertainties (5\%-–30\%) are much higher than would be naively expected from the S/N of the companion detection, as they include an estimate of the systematic errors resulting from the AO imaging, PSF-subtraction and contrast-measurement processes.

To obtain the separation and position angle of the binaries we measured the centroid of the PSF-subtracted images of the companion and primary, as above. We converted the raw pixel positions to on-sky separations and position angles using a distortion solution produced from Robo-AO measurements of globular clusters, detailed in \citet{Riddle15}. We calculated the uncertainties of the companion separation and position angles using estimated systematic errors in the position measurements due to blending between components, depending on the separation of the companion (typically 1-2 pixels uncertainty in the position of each star). We also included an estimate of the maximal changes in the Robo-AO orientation throughout the observation period ($\pm1\fdg5$), as verified using the globular cluster measurements above. Finally, we verified the measured positions and contrast ratios in direct measurement from non-PSF-subtracted images.

\section{NIRC2 data reduction}
\label{sec:kdata}

We created a pipeline to automatically reduce and analyze our NIRC2 data. After sky subtraction and flat-field calibration, the frames were co-added into a single image based on the automatic detection of the location of the primary star in each dither frame. Because the distortion across each detector quadrant is sufficiently small compared to the Robo-AO position errors, $\leq$20mas \citep{Yelda10}, we did not correct for field distortion. The pipeline then automatically identifies companion stars via pixel binning, vetting the brightest bins by measuring radius in the eight cardinal and diagonal directions against a minimum cutoff and radial consistency. Cutoff values are optimized to search for both wide and narrow separation companions. For targets with observations in multiple filters (reported in D. Atkinson et al., 2016, in preparation), the results are cross-referenced and stars found in only one filter are dropped as multi-color observations are effective in discriminating PSF speckles from astrophysical objects. Targets for which the pipeline registered multiple companions are flagged for manual validation.

The higher angular resolution images clearly separate the vast majority of companions and simple aperture photometry is used to measure the brightness of each star. For close companions, a matching aperture opposite the partner star from the object being investigated is subtracted from the star's own photometry. This corrects for any overlapping PSF halo (and assumes a radially constant PSF, which is more accurate at larger separations). Magnitude differences between companions and their primary star are shown in Table \ref{tab:table_keck_nirc2}. 

The uncertainty in the magnitude difference is estimated by varying the aperture radius from 0.5 to 5 of the width of the stellar PSF and measuring the standard deviation of the difference. Injected companions are used to measure the overall efficacy of the technique and determine the weighting of magnitude difference vs. aperture radius. Recovering injections with this technique demonstrated an uncertainty of $\sim5\%$, consistent with our reported uncertainties.

\section{Discoveries}
\label{sec:discoveries}

\begin{figure*}
\centering
\includegraphics[width=500pt]{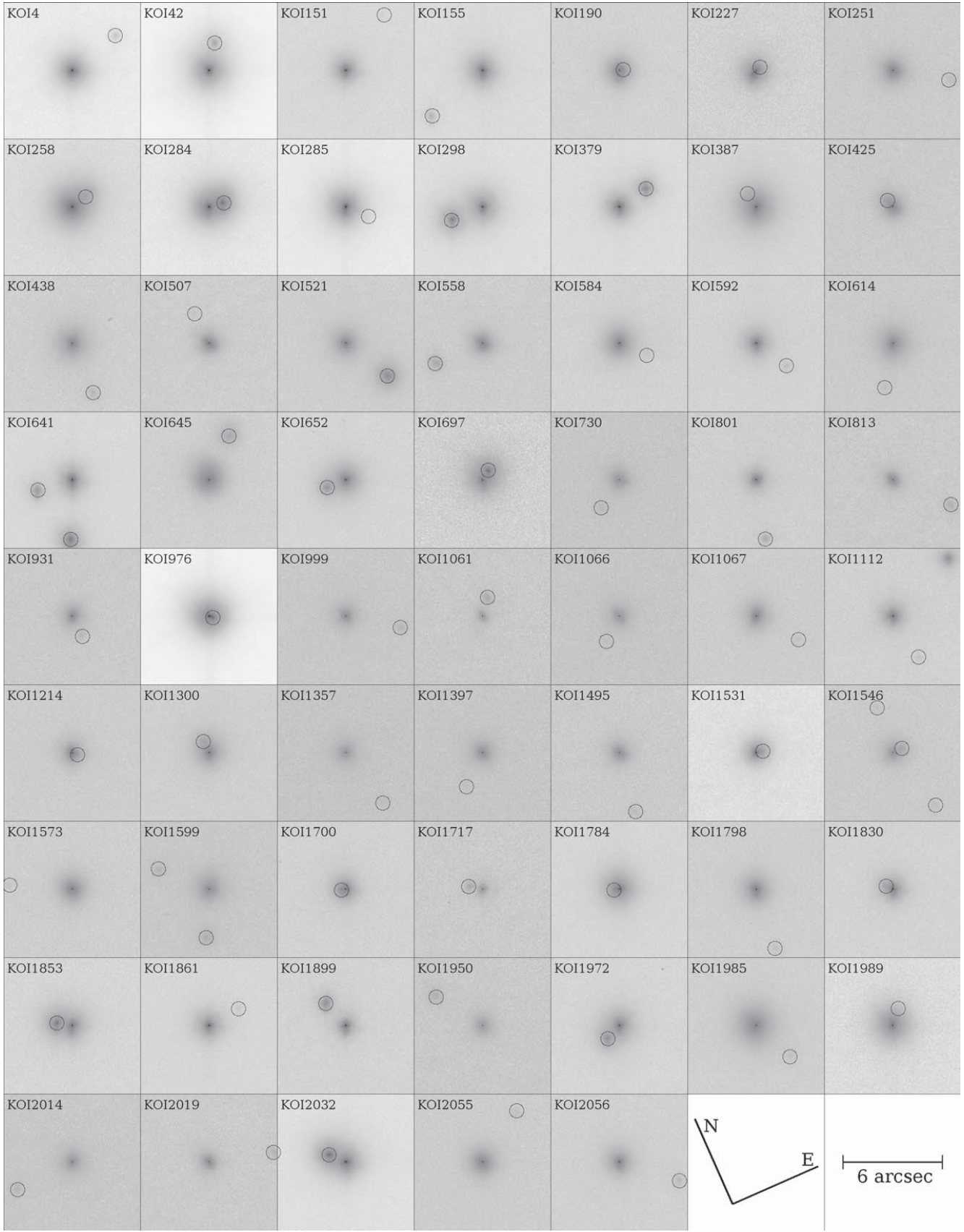}
\caption{Color inverted, normalized log-scale cutouts of 61 multiple KOI systems [KOI-4 to KOI-2056] with separations $<$4$\arcsec$ resolved with Robo-AO.  The angular scale and orientation is similar for each cutout.  The smaller circles are centered on the detected nearby star.}
\label{fig:cutouts1}
\end{figure*}

\begin{figure*}
\centering
\includegraphics[width=500pt]{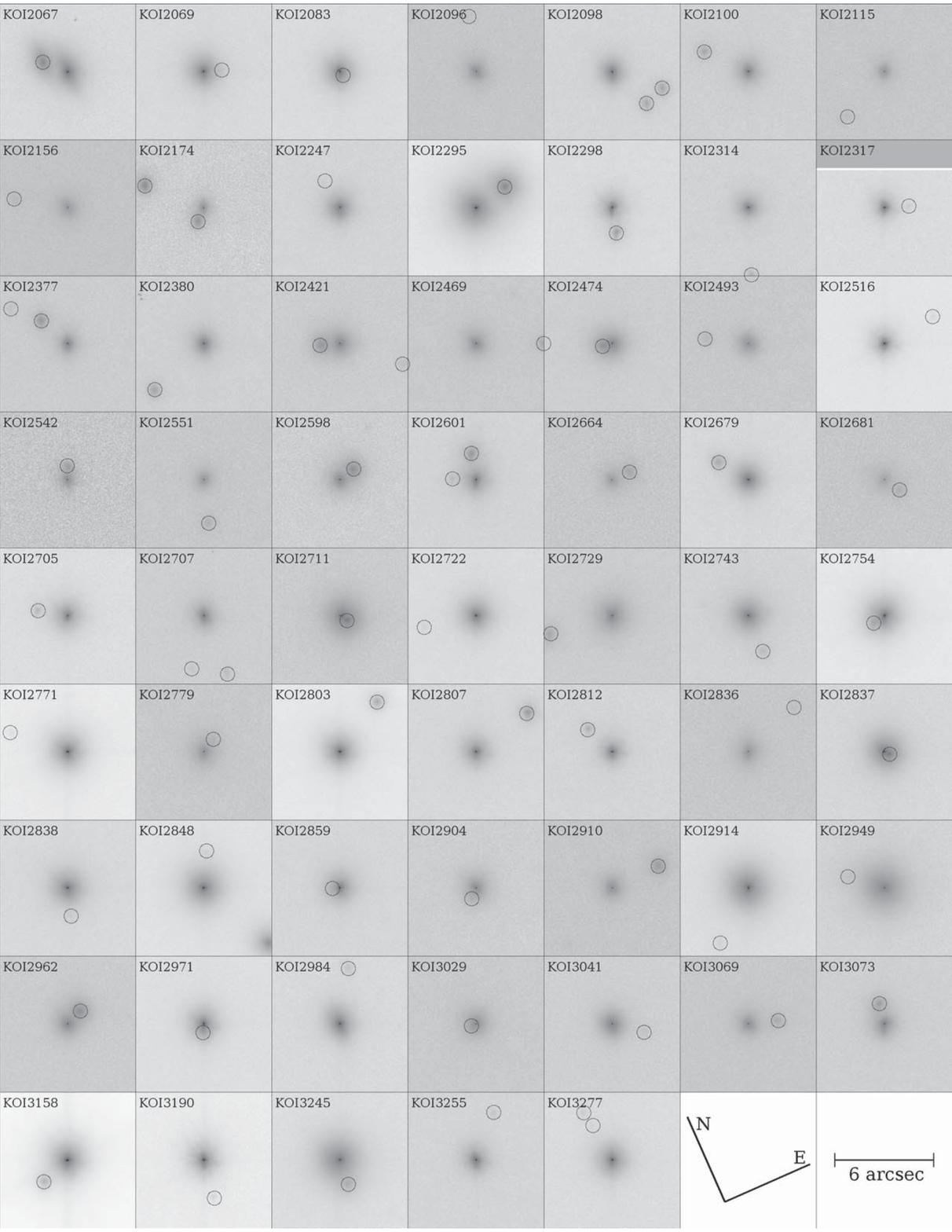}
\caption{Color inverted, normalized log-scale cutouts of 61 multiple KOI systems [KOI-2067 to KOI-3277] with separations $<$4$\arcsec$ resolved with Robo-AO.  The angular scale and orientation is similar for each cutout.  The smaller circles are centered on the detected nearby star.}
\label{fig:cutouts2}
\end{figure*}

\begin{figure*}
\centering
\includegraphics[width=500pt]{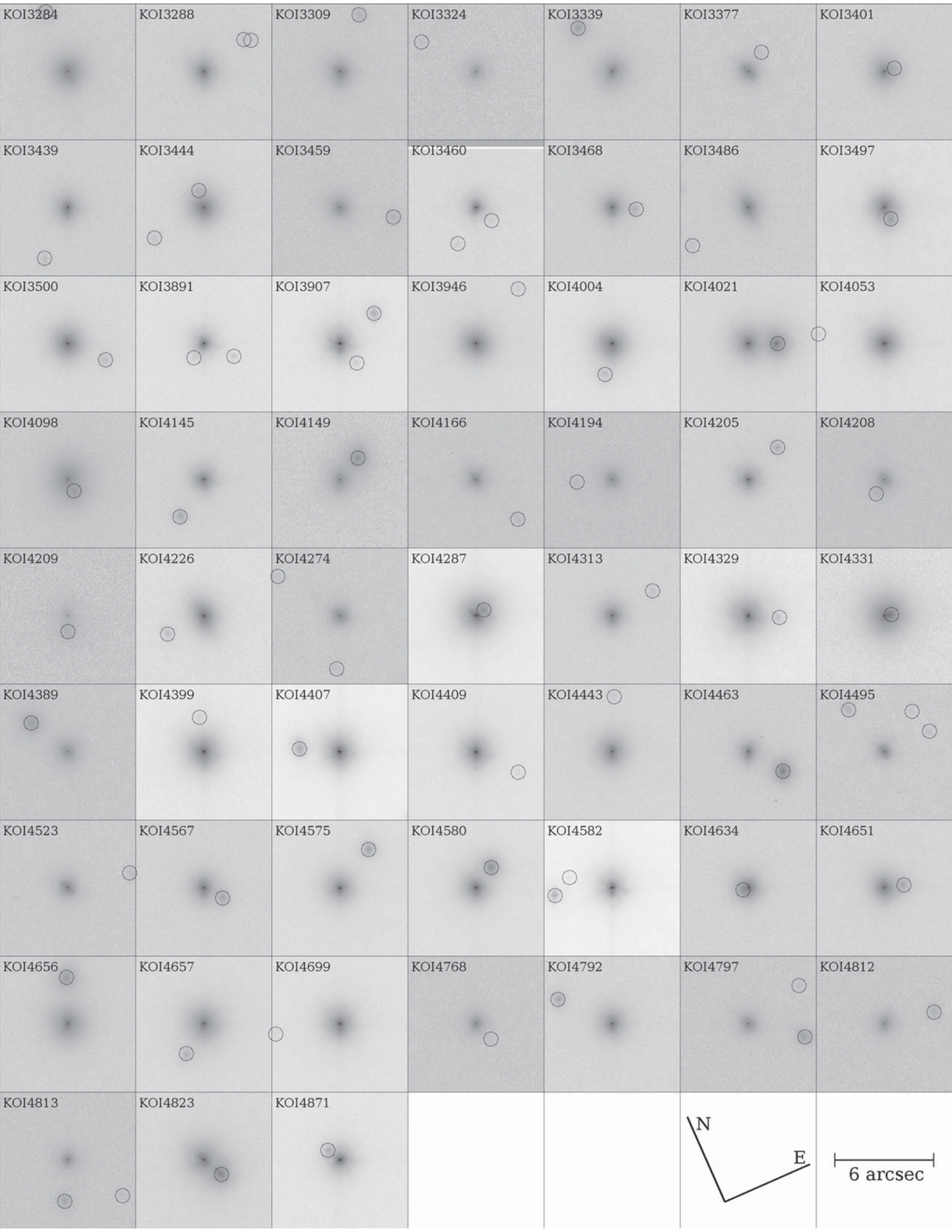}
\caption{Color inverted, normalized log-scale cutouts of 59 multiple KOI systems [KOI-3284 to KOI-4871] with separations $<$4$\arcsec$ resolved with Robo-AO.  The angular scale and orientation is similar for each cutout.  The smaller circles are centered on the detected nearby star.}
\label{fig:cutouts3}
\end{figure*}

We resolved 181 Kepler planet candidate hosts into multiple stars; the contrast ratios and the separations are shown in Figure \ref{fig:contrastcurves} and the discovery images are summarized in Figures \ref{fig:cutouts1}-\ref{fig:cutouts3}. The measured companion properties for the targets with secure detections, $>5\sigma$, are detailed in Table \ref{tab:comp_high}. Table \ref{tab:comp_low} describes probable companions which fell just below our formal $5\sigma$ detection criteria. We consider these very likely to be real, but we cannot exclude the possibility that a small fraction of these detections are spurious speckles without additional information. Where possible we observed these targets with NIRC2 to confirm their existence as discussed in Section \ref{sec:keck_add_obs}.

\subsection{Comparison to Other Imaging Surveys}
\label{sec:other_surveys}

In the search for blended \textit{Kepler} companions, many other high-angular-resolution surveys of KOIs have been performed using a range of observational techniques, e.g., infrared adaptive optics, sparse aperture interferometry, speckle interferometry, lucky imaging and direct imaging with HST \citep{A12, A13, D14, E15, G16, G15, H11, H12, K16, LB12, LB14, M14, Te15, To15, W15b, W15}. Each technique and instrument setup has a unique sensitivity, inner working angle, and spectral bandpass, presenting a challenge for a complete analysis of the multiplicity of KOI stars, and is part of our motivation for observing all KOIs with a single instrument setup. We have indicated in Tables \ref{tab:comp_high} and \ref{tab:comp_low} where these surveys have previously detected the same companion at approximately the same location. These other surveys have detected 38 of the 98 companions we detected at a significance greater than $5\sigma$, and 24 of the 105 companions we detected at a lower significance. 

Interestingly, Robo-AO was able to detect a close companion to KOI-2971 at a significance of $3.9\sigma$ that was not detected by \citet{D14} using ARIES with the MMT AO system; and conversely Robo-AO did not detect the 3\farcs5 companion found by \citet{D14}. We fortuitously observed KOI-2971 with NIRC2 (see Section \ref{sec:keck_add_obs}) and we were able to clearly see both of these companions.

We also note that the Robo-AO automated companion detection software found companions to KOI-2849 and KOI-3246 at nearly the same separation, 0\farcs36, position angle, $215^{\circ}$, magnitude difference, $\sim1$, and at a significance level of 4.0 $\sigma$ and 5 $\sigma$ respectively. Observations of both targets occurred on 2013 August 16 and showed evidence of static non-common path error in the stellar PSF leading to a speckle in the position of the purported companions. \citet{K16} observed KOI-3246 with NIRC2, and confirmed that there was no companion of similar brightness near the location of the Robo-AO automated detection. Because of our insensitivity at that position during that night we did not report these detections in our list of companions.

\subsection{Comparison to a Spectroscopic Survey}
\label{sec:spec_surveys}

\citet{K15} searched for close companions to KOIs by detecting secondary light sources in spectra used for determining radial velocities. They demonstrated sensitivity to companions as faint as $\sim$1\% of the primary star, and to companions captured within the 0\farcs87 $\times$ 3\farcs0 Keck-HIRES entrance slit. We observed 19 of the 58 KOIs for which \citet{K15} found evidence for a companion. Of those 19, we detected companions to KOI-151, KOI-652, KOI-1784 and KOI-4871. We had sufficient image contrast performance for the other 15 targets, 12 `high' and 3 `medium', to detect the brightness of companion indicated by \citet{K15} if they were separated by greater than roughly 0\farcs75; this lends evidence to the spectroscopic companions existing at closer angular separations.

We detected a companion to KOI-151 at an angular separation of 4\farcs2 and at a radiant flux ratio of 0.0046$\pm$0.0008 with respect to the primary star. This companion is fainter than the $1\sigma$ lower limit on the radiant flux of 0.012$\pm$0.006 determined by \citet{K15} and was likely not captured by the HIRES entrance slit. Therefore we conclude that our detection is new and that KOI-151 is an asterism of at least three stars. 

We detected a companion to KOI-652 at an angular separation of 1\farcs23. This companion was also detected by \citet{Te15} and \citet{K16} with Keck-NIRC2 as a $\sim$0\farcs08 binary. This lends further evidence to \citeauthor{K15}'s claim that this system is at least a triple star system.

Both we and \citet{W15} detected a companion to KOI-1784 at an angular separation of $\sim$0\farcs3, which would easily be captured within the HIRES entrance slit. We calculate a radiant flux ratio of 0.59$\pm$0.07 which is well above the lower limit of 0.192$\pm$0.058 determined by \citet{K15}. \citet{W15} estimated a physical separation of 160.7 AU between the primary and the companion, and this may not be compatible with the $\Delta$RV of -13 km s$^{-1}$ measured by \citet{K15}. It is not conclusive that the imaged close companion is also responsible for the spectroscopic signal, so it is possible this is also an asterism of at least three stars.

We observed a companion to KOI-4871 at an angular separation of 0\farcs96 and a radiant flux ratio of 0.057$\pm$0.010 with respect to the primary star. This detection is compatible with the lower limit on the radiant flux ratio of 0.012$\pm$0.003 determined by \citet{K15}. If this is indeed the same star, \citeauthor{K15}'s reported $\Delta$RV of -23 km s$^{-1}$ suggests it is not physically associated with the primary.  

While we have not conclusively imaged stellar companions that were detected by \citeauthor{K15}'s survey, we have found additional nearby stars not detected by spectroscopic methods. As previously suggested by \citet{Te15}, spectroscopic and AO methods can probe complementary, and sometimes overlapping, regions of parameter space when searching for stellar companions. While spectroscopic companions can be detected at much closer angular separations, AO observations probe larger angular separations, and at a much greater dynamic range that can be used to more precisely calibrate transit radii measurements. Physical association of companions can be established either from the difference in measured radial velocities from spectroscopy or from probabilistic or additional spectrophotometric parallax analysis when using AO. From a practical perspective, adaptive optics imaging requires much less on-sky observing time, and can target a much greater range of exoplanetary host stars.

\subsection{Keck-NIRC2 imaging of Robo-AO observed KOIs}
\label{sec:keck_add_obs}

Details of the 50 observations and 63 companions detected with NIRC2 appear in Table \ref{tab:table_keck_nirc2}. We observed 14 KOIs with companions detected at $>5\sigma$ so we could later calculate spectrophotometric parallax distances to determine the probability of physical association. We observed 13 companions detected by Paper I at a significance level of $<5\sigma$. Including those previously imaged with NIRC2 and by \citet{LB12, A12}, all 17 of these detections have been confirmed. 

The remainder of KOIs we observed with NIRC2 were at various levels of analysis: some Robo-AO images had been fully processed with measured significance levels on the candidate companion, others were manually identified before PSF subtraction. We confirmed 24 of the companions detected at a significance level of $<5\sigma$ in this survey (2 of which were in systems with a $>5\sigma$ detected companion), and other surveys confirmed 18 more of these companions. Due to the enhanced contrast with NIRC2, we found additional companions to 6 KOIs (1884, 2377, 2971, 3029, 3377 and 4407) that were not detected in the original Robo-AO data. We also found individual companions near KOI-2363 and KOI-4292 in the NIRC2 data that did not correspond to the preliminary manually-identified candidate companions in the Robo-AO data and we note them in Tables \ref{tab:table_keck_nirc2} and \ref{tab:obs_table}. Despite the somewhat haphazard selection of targets, the vast majority of Robo-AO detections that have follow-up observations with NIRC2 have been shown to be real; so far, 60 of the 120 companions detected at a significance of $<5\sigma$ with Robo-AO from Paper I and this work have been confirmed. 

\section{Discussion}
\label{sec:conclusions}

We observed 969 Kepler planetary system candidates with the Robo-AO robotic laser adaptive optics system. Presuming all confirmed and probable 203 Robo-AO detected companions within $\sim$4\arcsec of 181 KOIs are real, we calculate the nearby-star probability as a function of angular distance, see Figure \ref{fig:bin_frac_sep}. In our previous study we found a probability of 7.4\% $\pm$ 1.0\% at angular separations up to 2\farcs5 around 715 KOIs. For direct comparison, we calculated the probability in our sample to the same separation of 2\farcs5 and found it to be 10.6\% $\pm$ 1.1\%, a difference of 2.2 $\sigma$. Additional fainter companions are discovered here, e.g., 9 companions with a magnitude difference greater than 5 within 2\farcs5 compared to 3 previously.

\begin{figure*}
\centering
\includegraphics[width=0.83\paperwidth]{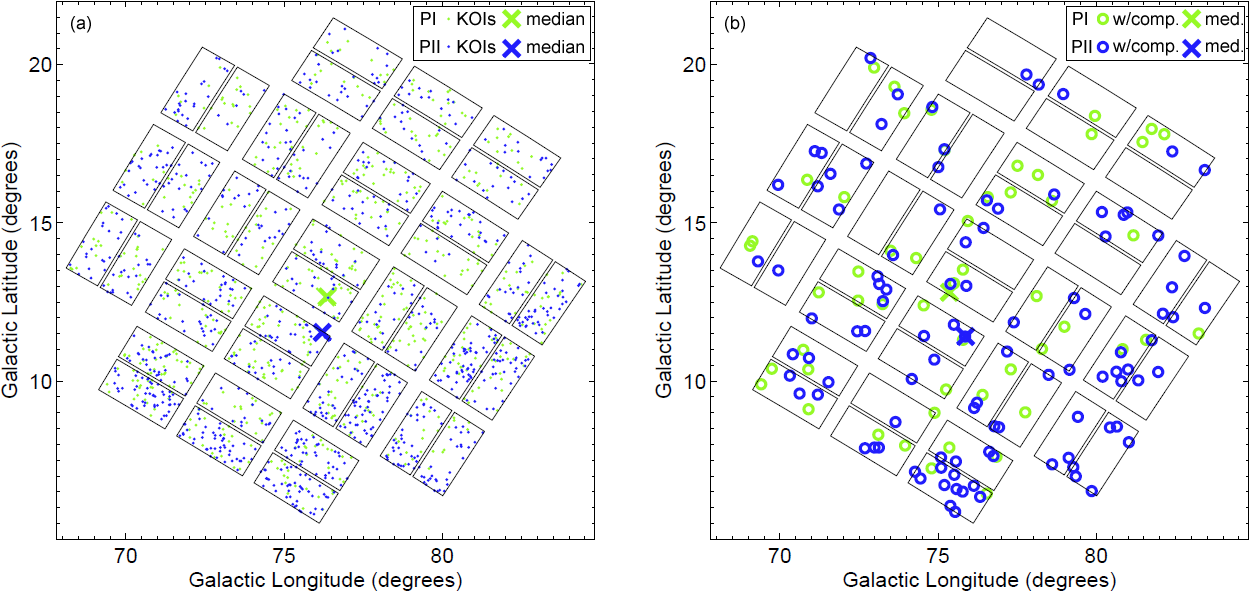}
\caption{Location in galactic coordinates for the KOIs (a), and those with companions detected at angular separations of less than 2\farcs5 (b), observed by Paper I (PI) and in this work (PII). In both figures the `$\times$' represents the median galactic coordinate for each set of objects. Projections of the \textit{Kepler} detectors on sky for the Spring season is provided for reference.}
\label{fig:field}
\end{figure*}

We explored the possibility of a bias change in the KOI selection process between major data releases, Q1-Q6 \citep{batalha13} and Q1-Q12 \citep{Rowe2015}. Our first study comprised 715 targets solely from Q1-Q6, while this work includes targets originally identified in both catalogs, 505 that appear in Q1-Q6 (51 of which have $<$2\farcs5 companions) and 464 that only appear in Q1-Q12 (52 companions). After combining the results of both and comparing the nearby-star probability within 2\farcs5 for the KOIs that only appear in Q1-Q12 catalog versus those originally found in the Q1-Q6 catalog, we find probabilities of 11.2\%$\pm1.6$\% and 8.5\%$\pm0.9$\% respectively, a difference of 1.5 $\sigma$.

We also examined the on-sky spatial distribution of observed KOIs and those with $<$2\farcs5 companions for both studies (Figure \ref{fig:field}). In our cumulative survey there is a bias of targets located closer to the galactic plane compared to the center of the \textit{Kepler} field. Additionally, compared to our first study, the median position of KOIs observed in this work is closer to the galactic plane by 1.1$^{\circ}$, and the median position of KOIs with $<$2\farcs5 companions is closer to the galactic plane by 1.4$^{\circ}$. We find in general that higher-galactic-latitude KOIs have fewer wide companions which is consistent with lower stellar crowding away from the galactic plane in the \textit{Kepler} field \citep{2011ApJS..197....6G}, and therefore the difference in nearby star probability between this work and Paper I may simply be due to the specific KOI samples.

Previous studies have shown that the majority of stellar companions to KOIs at a separation of $<$1\arcsec are physically associated, with the probability of association decreasing with increasing angular separation \citep{Horch2014}. We calculate the cumulative nearby-star probability as a function of angular separation of our data and present it in Figure \ref{fig:bin_frac_sep}. Our data show that the probability increases nearly linearly with increasing separation, up to $17.9$\%$\pm1.4$\% out to a separation of 4\farcs0. If the distribution had consisted solely of chance alignments of non-physically associated companions, we would expect this probability to instead increase quadratically. This suggests again that a large fraction of the companions with smaller angular separations are indeed physically associated with the primary star.

This work and Paper I together comprise a survey of roughly half of the KOIs in the Q1-Q17 DR24 dataset release. In light of the apparent discrepancy in companion discovery rates between the two using the same instrument, we caution against extrapolating companion rates from any individual survey that samples a small fraction of the overall population. Even when combined, the existing patchwork of other KOIs surveys (see Section \ref{sec:other_surveys}) is not as comprehensive, and requires detailed calibration to match the varying sensitivities, inner working angles and wavelength ranges. Future high-angular-resolution follow-up observations of large numbers of candidate exoplanet hosts would benefit from an initial comprehensive survey from Robo-AO that can very efficiently find lower contrast blended stars; preserving precious and limited resources like Keck AO or HST for those targets that pass the initial round of vetting.

\begin{figure}[b]
  \centering
  \resizebox{1.0\columnwidth}{!}
   {
    \includegraphics{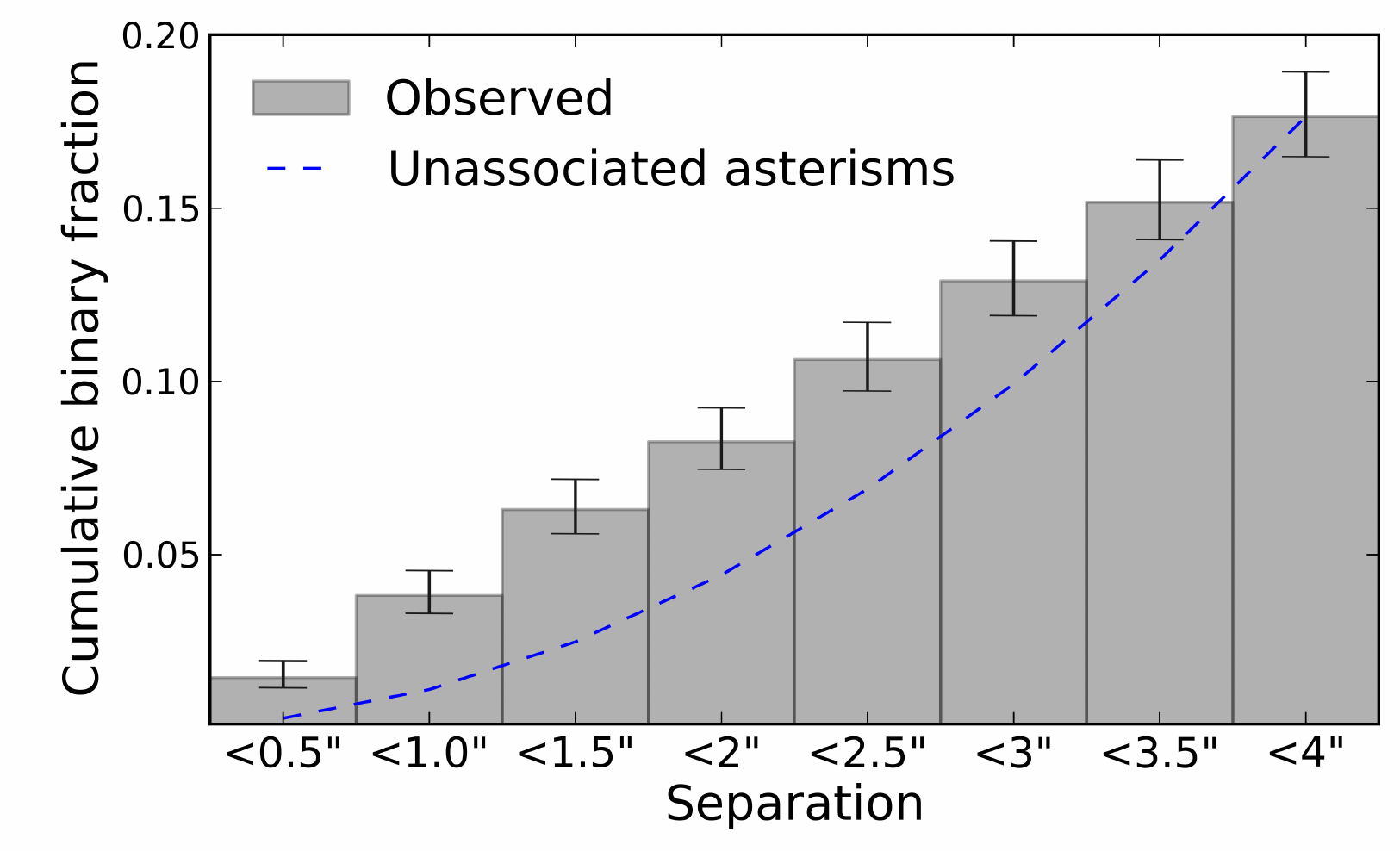}
   }
   \caption{Cumulative companion fraction for Robo-AO observed KOIs as a function of angular separation. The dashed line represents a theoretical quadratic cumulative distribution that would be expected from non-physically associated companions.\label{fig:bin_frac_sep}}

\end{figure}

We expect that our data will be used by other researchers using \textit{Kepler} data to study and validate exoplanets, their host stars and stellar environments, and other astrophysical phenomena. To aid in this effort, Robo-AO images of KOIs and the position and photometry of any detected companions will be available at the \textit{Kepler} Community Follow-up Observing Program\footnote{See https://exofop.ipac.caltech.edu/cfop.php}.

\section{Future Work}
\label{sec:future_work}

In the third installment of this paper series we will present the remaining Robo-AO observations of KOIs and explore an analysis of the complete data set. We will detail the effects of the detected nearby stars on the interpretation of \textit{Kepler} planetary candidates and note particular systems of interest. We will investigate further the spatial distribution of all KOIs with companions as a function of separation.

In parallel, we are currently using the multi-color visible and infrared observations obtained for this survey to estimate relative spectrophotometric distances between KOIs and their detected companions to determine if the stars are physically bound. We have also inspected all of our Robo-AO images of KOIs for images of other \textit{Kepler} input catalog (KIC) targets that do not have repeating transit signals. We have identified nearly 700 of these serendipitously observed KICs in our existing data; we will use these observations as a control sample to determine if there is a fundamental difference in the nearby-star probability between the KOIs and non-KOI KICs, including if there are similar effects due to galactic latitude, and whether this has an effect on planetary systems \citep{Nofi15}.

We have reconfigured and redeployed the Robo-AO system as the only instrument on the 2.1-m telescope at Kitt Peak as of November 2015 \citep{RAO_KP}. We intend to continue observations of our detected companions to search for common-proper-motion pairs to better understand the probability of physical association. We will additionally integrate a science-grade-detector version of a low-noise infrared camera to Robo-AO (previously demonstrated by \citealt{Baranec2015}). This camera will enable both infrared imaging and tip-tilt sensing and correction that will allow us to better observe redder KOIs. 

We are also in the process of building an upgraded Robo-AO system for the University of Hawai`i 2.2-m telescope on Maunakea \citep{Robo-AO2}. Between the two Robo-AO systems, we will be able observe up to $\sim$500 objects per night, covering nearly three-quarters of the sky over the course of a year. Forthcoming transit missions such as NASA's Transiting Exoplanet Survey Satellite \citep{TESS} scheduled to launch in 2017, and ESA's PLAnetary Transits and Oscillations of stars 2.0 \citep{PLATO} will release their data on timescales of months and shorter, and are expected to discover a greater number of exoplanet systems compared to \textit{Kepler}. Only the extremely efficient and rapid follow-up capability of Robo-AO will be able to keep up with the sustained demand for high-acuity imaging of thousands of exoplanet candidate host stars identified by these and other projects.

\acknowledgments

This research is supported by the NASA Exoplanets Research Program, grant $\#$NNX 15AC91G. C.B. acknowledges support from the Alfred P. Sloan Foundation. T.M. is supported by NASA grant $\#$NNX 14AE11G under the Kepler Participating Scientist Program. D.A. is supported by a NASA Space Technology Research Fellowship, grant $\#$NNX 13AL75H.

The Robo-AO system was developed by collaborating partner institutions, the California Institute of Technology and the Inter-University Centre for Astronomy and Astrophysics, and with the support of the National Science Foundation under Grant Nos. AST-0906060, AST-0960343, and AST-1207891, the Mt. Cuba Astronomical Foundation and by a gift from Samuel Oschin. We are grateful to the Palomar Observatory staff for their support of Robo-AO on the 1.5 m telescope, particularly S. Kunsman, M. Doyle, J. Henning, R. Walters, G. Van Idsinga, B. Baker, K. Dunscombe and D. Roderick. We thank Adam Kraus et al. for sharing a preprint of their paper. This work used the astronomy \& astrophysics package for Matlab \citep{Ofek14}. Some of the data presented herein were obtained at the W.M. Keck Observatory, which is operated as a scientific partnership among the California Institute of Technology, the University of California and the National Aeronautics and Space Administration. The Observatory was made possible by the generous financial support of the W.M. Keck Foundation. The authors wish to recognize and acknowledge the very significant cultural role and reverence that the summit of Maunakea has always had within the indigenous Hawaiian community.  We are most fortunate to have the opportunity to conduct observations from this mountain.



{\it Facilities:} \facility{PO:1.5m (Robo-AO), Keck:II (NIRC2-LGS).}

\section*{Appendix}

In Table \ref{tab:obs_table}, we list our Robo-AO observed KOIs, including date the target was observed, the observation quality as described in Section \ref{sec:imageperf}, and the presence of detected companions.

\bibliographystyle{apj.bst}   
\bibliography{references.bbl}   




\clearpage

\clearpage

\LongTables 


\clearpage

\begin{deluxetable*}{lcccccrcll}
\tablewidth{0pt}
\tablecaption{\label{tab:comp_high}Detections of Objects within $\sim$4\farcs0 of \textit{Kepler} Planet Candidates at $\geq5\sigma$ Significance}
\tablehead{
\colhead{KOI} & \colhead{$m_{i}$} &\colhead{ObsID}  & \colhead{Filter}  & \colhead{Signf.} & \colhead{Separation} & \colhead{P.A.} & \colhead{Mag. Diff.} & \colhead{Previous} & \colhead{NIRC2} \\
\colhead{} & \colhead{(mag)} &\colhead{}  & \colhead{}  & \colhead{$\sigma$} & \colhead{(arcsec)} & \colhead{(deg.)} & \colhead{(mag)} & \colhead{Detection?} & \colhead{Detection?} \\
}
\startdata
KOI-4 & 11.3 & 2012/07/16 & $i'$ & 12 & 3.42$\pm$0.06 & 75$\pm$2 & 4.46$\pm$0.05 &  & \\
KOI-42 & 9.2 & 2013/07/28 & LP600 & 35 & 1.74$\pm$0.06 & 35$\pm$2 & 3.04$\pm$0.17 & A12, H11, K16 & \\ 
KOI-227 & 13.7 & 2013/07/27 & LP600 & 5 & 0.33$\pm$0.06 & 72$\pm$6 & 0.84$\pm$0.09 & H11, K16 & \\ 
KOI-258 & 9.8 & 2012/07/18 & $i'$ & 14 & 1.05$\pm$0.06 & 77$\pm$2 & 2.76$\pm$0.17 & A12, H11 & \\ 
KOI-284 & 11.7 & 2013/07/27 & LP600 & 19 & 0.96$\pm$0.06 & 98$\pm$2 & 0.45$\pm$0.04 & A12, H11, E15, K16 & \\ 
KOI-298 & 12.4 & 2013/08/15 & LP600 & 477 & 2.11$\pm$0.06 & 270$\pm$2 & 0.58$\pm$0.04 & LB12, K16 & \\ 
KOI-379 & 13.2 & 2013/07/27 & LP600 & 56 & 2.11$\pm$0.06 & 83$\pm$2 & 1.42$\pm$0.11 & LB12 & \\ 
KOI-521 & 14.5 & 2013/08/14 & LP600 & 18 & 3.24$\pm$0.06 & 152$\pm$2 & 0.42$\pm$0.03 &  & \\ 
KOI-558 & 14.6 & 2013/08/22 & LP600 & 6 & 3.16$\pm$0.06 & 271$\pm$2 & 2.06$\pm$0.04 &  & \\ 
KOI-641 & 13.1 & 2013/07/24 & LP600 & 18 & 2.09$\pm$0.06 & 278$\pm$2 & 2.07$\pm$0.05 & LB12 & \\ 
 &  &  &  & 46 & 3.65$\pm$0.06 & 205$\pm$2 & 0.33$\pm$0.06 & LB12 & \\ 
KOI-645 & 13.5 & 2013/07/25 & LP600 & 6 & 2.98$\pm$0.06 & 48$\pm$2 & 2.23$\pm$0.04 & LB12 & \\ 
KOI-652 & 13.3 & 2013/07/28 & LP600 & 21 & 1.23$\pm$0.06 & 272$\pm$2 & 1.59$\pm$0.14 & K15, K16, Te15 & \\ 
KOI-697 & 13.5 & 2013/08/15 & LP600 & 9 & 0.71$\pm$0.06 & 54$\pm$3 & 0.06$\pm$0.03 & W15 & \\ 
KOI-801 & 14.8 & 2013/10/25 & LP600 & 6 & 3.67$\pm$0.06 & 195$\pm$2 & 2.58$\pm$0.11 &  & \\ 
KOI-976 &  & 2012/08/03 & $i'$ & 22 & 0.25$\pm$0.06 & 129$\pm$7 & 0.34$\pm$0.09 & K16 & \\ 
KOI-1061 & 14.3 & 2013/08/20 & LP600 & 14 & 1.22$\pm$0.06 & 38$\pm$2 & 1.21$\pm$0.07 &  & \\ 
KOI-1300 & 13.9 & 2013/08/16 & LP600 & 8 & 0.78$\pm$0.06 & 357$\pm$3 & 1.79$\pm$0.18 & K16 & \\ 
KOI-1357 & 15.3 & 2013/08/18 & LP600 & 30 & 3.83$\pm$0.06 & 167$\pm$2 & 3.38$\pm$0.03 &  & \\ 
KOI-1531 & 12.9 & 2013/08/16 & LP600 & 7 & 0.43$\pm$0.06 & 99$\pm$4 & 0.90$\pm$0.16 &  & \\ 
KOI-1546 & 14.2 & 2013/08/19 & LP600 & 6 & 0.62$\pm$0.06 & 86$\pm$3 & 1.03$\pm$0.12 & LB14 & yes \\ 
KOI-1717 & 14.3 & 2013/10/25 & LP600 & 9 & 0.87$\pm$0.06 & 305$\pm$3 & 1.46$\pm$0.13 &  & \\ 
KOI-1853 & 13.3 & 2013/08/14 & LP600 & 11 & 0.96$\pm$0.06 & 304$\pm$2 & 0.24$\pm$0.05 &  & \\ 
KOI-1861 & 13.8 & 2013/07/28 & LP600 & 5 & 2.10$\pm$0.06 & 84$\pm$2 & 4.93$\pm$0.16 &  & \\ 
KOI-1899 & 14.4 & 2013/07/27 & LP600 & 16 & 1.84$\pm$0.06 & 342$\pm$2 & 0.94$\pm$0.05 &  & \\ 
KOI-1972 & 13.6 & 2013/08/15 & LP600 & 21 & 1.05$\pm$0.06 & 246$\pm$2 & 1.05$\pm$0.12 &  & \\ 
KOI-2032 & 12.0 & 2013/07/28 & LP600 & 17 & 1.19$\pm$0.06 & 317$\pm$2 & 0.34$\pm$0.05 & K16 & \\ 
KOI-2067 & 12.3 & 2013/07/24 & LP600 & 17 & 1.64$\pm$0.06 & 315$\pm$2 & 0.80$\pm$0.04 & K16 & \\ 
KOI-2096 & 14.9 & 2013/08/19 & LP600 & 29 & 3.50$\pm$0.06 & 17$\pm$2 & 4.13$\pm$0.11 &  & \\ 
KOI-2098 & 13.7 & 2013/07/27 & LP600 & 8 & 2.88$\pm$0.06 & 156$\pm$2 & 2.58$\pm$0.04 &  & \\ 
 &  &  &  & 10 & 3.24$\pm$0.06 & 132$\pm$2 & 2.40$\pm$0.06 &  & \\ 
KOI-2100 & 14.4 & 2013/07/28 & LP600 & 8 & 2.98$\pm$0.06 & 318$\pm$2 & 2.10$\pm$0.05 &  & \\ 
KOI-2174 & 15.2 & 2012/08/07 & LP600 & 10 & 0.92$\pm$0.06 & 226$\pm$2 & 0.21$\pm$0.06 &  & \\ 
 &  &  &  & 25 & 3.88$\pm$0.06 & 314$\pm$2 & 0.14$\pm$0.03 &  & \\ 
KOI-2295 & 11.4 & 2013/07/29 & LP600 & 21 & 2.19$\pm$0.06 & 78$\pm$2 & 0.88$\pm$0.06 & K16 & \\ 
KOI-2298 & 13.5 & 2013/08/16 & LP600 & 29 & 1.57$\pm$0.06 & 194$\pm$2 & 2.08$\pm$0.14 & D14 & \\ 
KOI-2377 & 14.5 & 2013/07/24 & LP600 & 14 & 2.09$\pm$0.06 & 335$\pm$2 & 1.25$\pm$0.03 &  & yes \\ 
KOI-2421 & 14.0 & 2013/07/25 & LP600 & 8 & 1.23$\pm$0.06 & 290$\pm$2 & 0.99$\pm$0.09 & D14 & \\ 
KOI-2474 & 13.9 & 2013/08/13 & LP600 & 8 & 0.61$\pm$0.06 & 279$\pm$3 & 0.65$\pm$0.07 & H12 & \\ 
KOI-2598 & 14.0 & 2013/08/13 & LP600 & 9 & 1.09$\pm$0.06 & 75$\pm$2 & 0.37$\pm$0.04 &  & \\ 
KOI-2601 & 13.8 & 2013/08/13 & LP600 & 17 & 1.66$\pm$0.06 & 14$\pm$2 & 1.43$\pm$0.06 &  & \\ 
KOI-2679 & 13.3 & 2013/07/25 & LP600 & 19 & 2.11$\pm$0.06 & 324$\pm$2 & 2.87$\pm$0.06 &  & \\ 
KOI-2705 & 14.3 & 2013/07/25 & LP600 & 15 & 1.84$\pm$0.06 & 304$\pm$2 & 3.19$\pm$0.14 & G16, K16 & yes\\ 
KOI-2711 & 13.5 & 2013/07/29 & LP600 & 8 & 0.52$\pm$0.06 & 147$\pm$4 & 0.12$\pm$0.08 &  & yes\\ 
KOI-2729 & 13.7 & 2013/08/18 & LP600 & 6 & 3.94$\pm$0.06 & 278$\pm$2 & 2.03$\pm$0.12 &  & \\ 
KOI-2754 &  & 2013/08/14 & LP600 & 21 & 0.79$\pm$0.06 & 260$\pm$3 & 2.23$\pm$0.20 & D14, K16 & \\ 
KOI-2771 &  & 2013/08/16 & LP600 & 13 & 3.85$\pm$0.06 & 312$\pm$2 & 6.61$\pm$0.14 & D14 & \\
KOI-2803 & 12.1 & 2013/08/13 & LP600 & 25 & 3.84$\pm$0.06 & 61$\pm$2 & 3.00$\pm$0.05 & D14, K16 & \\ 
KOI-2807 & 13.7 & 2013/07/28 & LP600 & 13 & 3.93$\pm$0.06 & 77$\pm$2 & 1.90$\pm$0.04 &  & \\ 
KOI-2812 & 14.2 & 2013/07/27 & LP600 & 22 & 2.09$\pm$0.06 & 335$\pm$2 & 3.23$\pm$0.05 &  & \\ 
KOI-2836 & 14.9 & 2013/10/22 & LP600 & 217 & 3.94$\pm$0.06 & 70$\pm$2 & 3.39$\pm$0.04 &  & \\ 
KOI-2837 & 13.1 & 2013/08/15 & LP600 & 6 & 0.35$\pm$0.06 & 136$\pm$5 & 0.23$\pm$0.04 &  & yes\\ 
KOI-2848 & 12.3 & 2013/08/14 & LP600 & 6 & 2.30$\pm$0.06 & 28$\pm$2 & 5.63$\pm$0.23 &  & \\ 
KOI-2904 & 12.5 & 2013/07/24 & LP600 & 10 & 0.71$\pm$0.06 & 226$\pm$3 & 1.99$\pm$0.24 & D14 & yes\\ 
KOI-2910 & 15.0 & 2013/08/19 & LP600 & 7 & 3.15$\pm$0.06 & 88$\pm$2 & 0.72$\pm$0.05 &  & \\ 
KOI-2962 & 14.0 & 2013/07/25 & LP600 & 9 & 1.13$\pm$0.06 & 68$\pm$2 & 0.53$\pm$0.05 &  & \\ 
KOI-3069 & 14.7 & 2013/08/19 & LP600 & 5 & 1.93$\pm$0.06 & 109$\pm$2 & 2.20$\pm$0.07 &  & yes\\ 
KOI-3073 & 14.2 & 2013/08/14 & LP600 & 10 & 1.30$\pm$0.06 & 10$\pm$2 & 1.76$\pm$0.15 &  & \\ 
KOI-3158 &  & 2013/07/21 & LP600 & 613 & 2.10$\pm$0.06 & 254$\pm$2 & 4.00$\pm$0.15 & LB14, K16, C15 & \\ 
KOI-3190 & 11.1 & 2013/07/27 & LP600 & 6 & 2.68$\pm$0.06 & 190$\pm$2 & 5.92$\pm$0.17 &  & \\ 
KOI-3245 & 12.3 & 2013/07/25 & LP600 & 12 & 1.58$\pm$0.06 & 184$\pm$2 & 3.10$\pm$0.07 &  & \\ 
KOI-3324 & 15.7 & 2013/10/23 & LP600 & 11 & 3.84$\pm$0.06 & 323$\pm$2 & 3.05$\pm$0.06 &  & \\ 
KOI-3339 & 14.4 & 2013/10/22 & LP600 & 9 & 3.41$\pm$0.06 & 346$\pm$2 & 1.38$\pm$0.06 &  & \\ 
KOI-3468 & 14.1 & 2013/07/24 & LP600 & 6 & 1.49$\pm$0.06 & 117$\pm$2 & 3.22$\pm$0.15 &  & \\ 
KOI-3497 & 13.0 & 2013/10/24 & LP600 & 15 & 0.78$\pm$0.06 & 174$\pm$3 & 1.23$\pm$0.12 & M14, K16 & \\ 
KOI-3891 & 13.4 & 2013/07/27 & LP600 & 5 & 1.05$\pm$0.06 & 240$\pm$2 & 4.69$\pm$0.41 & K16 & \\ 
 &  &  &  & 9 & 2.01$\pm$0.06 & 136$\pm$2 & 4.92$\pm$0.13 & K16 & \\ 
KOI-3907 & 12.5 & 2013/07/27 & LP600 & 7 & 1.58$\pm$0.06 & 162$\pm$2 & 6.31$\pm$0.25 & W15 & \\ 
 &  &  &  & 20 & 2.82$\pm$0.06 & 72$\pm$2 & 3.23$\pm$0.04 & W15 & \\ 
KOI-4004 & 12.5 & 2013/07/27 & LP600 & 13 & 1.93$\pm$0.06 & 217$\pm$2 & 4.34$\pm$0.27 & K16 & yes \\ 
KOI-4021 & 12.5 & 2013/08/16 & LP600 & 11 & 1.92$\pm$0.06 & 113$\pm$2 & 0.52$\pm$0.07 &  & \\ 
KOI-4145 & 14.1 & 2013/07/27 & LP600 & 9 & 2.71$\pm$0.06 & 237$\pm$2 & 2.36$\pm$0.03 &  & \\ 
KOI-4149 & 14.1 & 2013/10/22 & LP600 & 11 & 1.76$\pm$0.06 & 63$\pm$2 & 0.17$\pm$0.03 &  & \\ 
KOI-4205 & 14.2 & 2013/07/28 & LP600 & 7 & 2.71$\pm$0.06 & 66$\pm$2 & 2.65$\pm$0.03 &  & \\ 
KOI-4209 & 15.7 & 2013/08/19 & LP600 & 11 & 0.96$\pm$0.06 & 203$\pm$2 & 0.37$\pm$0.08 &  & yes\\ 
KOI-4287 & 11.1 & 2013/08/17 & LP600 & 10 & 0.61$\pm$0.06 & 76$\pm$3 & 1.27$\pm$0.14 & K16 & \\ 
KOI-4329 & 11.9 & 2013/08/18 & LP600 & 10 & 1.93$\pm$0.06 & 117$\pm$2 & 4.64$\pm$0.21 &  & \\ 
KOI-4331 & 13.0 & 2013/07/29 & LP600 & 6 & 0.45$\pm$0.06 & 103$\pm$4 & 0.25$\pm$0.04 &  & yes \\ 
KOI-4389 & 14.8 & 2013/08/21 & LP600 & 7 & 2.88$\pm$0.06 & 332$\pm$2 & 0.58$\pm$0.06 &  & \\ 
KOI-4399 & 11.8 & 2013/07/24 & LP600 & 6 & 2.16$\pm$0.06 & 17$\pm$2 & 6.24$\pm$0.21 & K16 & \\ 
KOI-4407 & 11.0 & 2013/07/24 & LP600 & 19 & 2.54$\pm$0.06 & 298$\pm$2 & 2.97$\pm$0.05 & E15, K16 & yes\\ 
KOI-4463 & 14.6 & 2013/07/27 & LP600 & 29 & 2.45$\pm$0.06 & 143$\pm$2 & 0.01$\pm$0.03 &  & yes\\ 
KOI-4495 & 15.2 & 2013/08/22 & LP600 & 6 & 3.06$\pm$0.06 & 89$\pm$2 & 3.90$\pm$0.06 &  & \\ 
 &  &  &  & 9 & 3.41$\pm$0.06 & 344$\pm$2 & 2.68$\pm$0.05 &  & \\ 
KOI-4567 & 13.6 & 2013/07/24 & LP600 & 12 & 1.31$\pm$0.06 & 142$\pm$2 & 2.48$\pm$0.13 &  & \\ 
KOI-4575 & 13.0 & 2013/08/16 & LP600 & 18 & 2.97$\pm$0.06 & 61$\pm$2 & 2.18$\pm$0.03 &  & \\ 
KOI-4580 & 12.8 & 2013/08/13 & LP600 & 19 & 1.58$\pm$0.06 & 60$\pm$2 & 1.27$\pm$0.05 &  & \\ 
KOI-4582 & 11.6 & 2013/07/27 & LP600 & 68 & 2.71$\pm$0.06 & 308$\pm$2 & 6.28$\pm$0.18 & K16 & \\ 
 &  &  &  & 39 & 3.55$\pm$0.06 & 286$\pm$2 & 3.27$\pm$0.12 & K16 & \\ 
KOI-4634 & 13.5 & 2013/07/24 & LP600 & 5 & 0.35$\pm$0.06 & 275$\pm$5 & 1.55$\pm$0.18 &  & yes\\ 
KOI-4651 & 13.6 & 2013/07/25 & LP600 & 9 & 1.22$\pm$0.06 & 105$\pm$2 & 2.88$\pm$0.37 &  & \\ 
KOI-4656 & 13.7 & 2013/07/29 & LP600 & 8 & 2.89$\pm$0.06 & 23$\pm$2 & 1.42$\pm$0.03 &  & \\ 
KOI-4657 & 13.0 & 2013/07/29 & LP600 & 22 & 2.11$\pm$0.06 & 234$\pm$2 & 3.27$\pm$0.10 & K16 & \\ 
KOI-4792 & 14.0 & 2013/08/15 & LP600 & 9 & 3.68$\pm$0.06 & 318$\pm$2 & 2.36$\pm$0.04 &  & \\ 
KOI-4797 & 15.3 & 2013/08/22 & LP600 & 6 & 3.59$\pm$0.06 & 127$\pm$2 & 1.12$\pm$0.06 &  & \\ 
KOI-4813 & 13.3 & 2013/07/28 & LP600 & 5 & 2.54$\pm$0.06 & 208$\pm$2 & 1.22$\pm$0.05 &  & \\ 
KOI-4823 & 12.5 & 2013/08/17 & LP600 & 16 & 1.40$\pm$0.06 & 153$\pm$2 & 0.59$\pm$0.04 &  & \\ 
KOI-4871 & 12.9 & 2013/10/25 & LP600 & 11 & 0.96$\pm$0.06 & 333$\pm$2 & 3.12$\pm$0.19 & $^{a}$ & yes\\ 
\enddata
\tablenotetext{}{\textbf{Notes:} References for previous detections are denoted with the following codes: \citealt{A12} (A12); \citealt{C15} (C15); \citealt{D14} (D14); \citealt{E15} (E15); \citealt{G16} (G16); \citealt{H11} (H11); \citealt{H12} (H12); \citealt{K16} (K16); \citealt{LB12} (LB12); \citealt{LB14} (LB14); \citealt{Mu14} (M14); \citealt{Te15} (Te15); \citealt{W15} (W15).}
\tablenotetext{a}{Companion identity is ambiguous. See Section \ref{sec:spec_surveys}.}
\end{deluxetable*}

\clearpage

\begin{deluxetable*}{lcccccrcll}
\tablewidth{0pt}
\tablecaption{\label{tab:comp_low}Detections of Objects within $\sim$4\farcs0 of \textit{Kepler} Planet Candidates at $<5\sigma$ Significance}
\tablehead{
\colhead{KOI} & \colhead{$m_{i}$} &\colhead{ObsID}  & \colhead{Filter}  & \colhead{Signf.} & \colhead{Separation} & \colhead{P.A.} & \colhead{Mag. Diff.} & \colhead{Previous} & \colhead{NIRC2} \\
\colhead{} & \colhead{(mag)} &\colhead{}  & \colhead{}  & \colhead{$\sigma$} & \colhead{(arcsec)} & \colhead{(deg.)} & \colhead{(mag)} & \colhead{Detection?} & \colhead{Detection?} \\
}
\startdata
KOI-151 & 13.8 & 2013/07/27 & LP600 & 4.2 & 4.17$\pm$0.06 & 58$\pm$2 & 5.84$\pm$0.18 &   & \\ 
KOI-155 & 13.3 & 2013/08/13 & LP600 & 4.0 & 4.01$\pm$0.06 & 251$\pm$2 & 3.83$\pm$0.17 &   & \\ 
KOI-190 & 13.9 & 2013/07/27 & LP600 & 3.6 & 0.23$\pm$0.06 & 105$\pm$2 & 1.33$\pm$0.18 &   & yes\\ 
KOI-251 & 14.1 & 2013/08/21 & LP600 & 2.1 & 3.48$\pm$0.06 & 123$\pm$2 & 3.80$\pm$0.12 & A12, G16, K16  & \\ 
KOI-285 &  & 2013/08/13 & LP600 & 4.6 & 1.51$\pm$0.06 & 136$\pm$2 & 6.12$\pm$0.21 & A12, K16  & \\ 
KOI-387 & 13.2 & 2013/07/29 & LP600 & 2.3 & 0.98$\pm$0.06 & 352$\pm$2 & 3.86$\pm$0.18 & LB12, K16  & \\ 
KOI-425 & 14.5 & 2013/08/22 & LP600 & 2.8 & 0.53$\pm$0.06 & 346$\pm$4 & 0.86$\pm$0.10 &   & yes\\ 
KOI-438 & 13.8 & 2013/08/18 & LP600 & 3.9 & 3.28$\pm$0.06 & 181$\pm$2 & 3.11$\pm$0.04 & K16  & \\ 
KOI-507 & 14.6 & 2013/08/22 & LP600 & 2.3 & 2.03$\pm$0.06 & 358$\pm$2 & 4.46$\pm$0.11 &   & \\ 
KOI-584 & 13.9 & 2012/08/05 & LP600 & 2.5 & 1.83$\pm$0.06 & 137$\pm$2 & 4.10$\pm$0.12 &   & \\ 
KOI-592 & 14.1 & 2013/08/14 & LP600 & 2.6 & 2.30$\pm$0.06 & 150$\pm$2 & 4.21$\pm$0.11 & LB12  & \\ 
KOI-614 & 14.3 & 2013/07/29 & LP600 & 2.2 & 2.76$\pm$0.06 & 214$\pm$2 & 4.01$\pm$0.03 &   & \\ 
KOI-730 & 15.1 & 2013/08/21 & LP600 & 3.8 & 2.04$\pm$0.06 & 237$\pm$2 & 2.95$\pm$0.09 &   & \\ 
KOI-813 & 15.5 & 2013/08/22 & LP600 & 4.3 & 3.87$\pm$0.06 & 137$\pm$2 & 2.09$\pm$0.08 &   & \\ 
KOI-931 & 15.0 & 2013/10/23 & LP600 & 4.0 & 1.38$\pm$0.06 & 177$\pm$2 & 3.40$\pm$0.14 &   & yes\\ 
KOI-999 & 15.0 & 2013/08/18 & LP600 & 2.6 & 3.41$\pm$0.06 & 125$\pm$2 & 2.80$\pm$0.05 &   & \\ 
KOI-1066 & 15.4 & 2013/08/21 & LP600 & 2.2 & 1.69$\pm$0.06 & 205$\pm$2 & 4.19$\pm$0.18 &   & yes\\ 
KOI-1067 & 14.5 & 2013/10/23 & LP600 & 3.5 & 2.97$\pm$0.06 & 143$\pm$2 & 4.05$\pm$0.15 &   & yes\\ 
KOI-1112 & 14.4 & 2013/07/27 & LP600 & 2.5 & 2.95$\pm$0.06 & 172$\pm$2 & 4.57$\pm$0.05 &   & yes\\
KOI-1214 & 14.4 & 2013/07/24 & LP600 & 3.9 & 0.33$\pm$0.06 & 132$\pm$2 & 1.21$\pm$0.18 &   & yes\\ 
KOI-1397 & 14.8 & 2013/08/21 & LP600 & 3.1 & 2.30$\pm$0.06 & 229$\pm$2 & 4.41$\pm$0.17 & K16  & \\ 
KOI-1495 & 15.2 & 2013/08/22 & LP600 & 2.5 & 3.75$\pm$0.06 & 188$\pm$2 & 2.92$\pm$0.11 &   & \\ 
KOI-1546 & 14.2 & 2013/08/19 & LP600 & 4.2 & 4.15$\pm$0.06 & 165$\pm$2 & 3.34$\pm$0.07 & LB14 & yes \\ 
 &  &  &  & 3.5 & 2.93$\pm$0.06 & 5$\pm$2 & 3.52$\pm$0.08 & LB14 & yes \\ 
KOI-1573 & 14.2 & 2013/07/25 & LP600 & 3.7 & 3.84$\pm$0.06 & 299$\pm$2 & 4.72$\pm$0.15 &   & \\ 
KOI-1599 & 14.6 & 2013/08/18 & LP600 & 3.4 & 2.98$\pm$0.06 & 207$\pm$2 & 2.22$\pm$0.06 &   & \\ 
 &  &  &  & 2.6 & 3.42$\pm$0.06 & 316$\pm$2 & 2.89$\pm$0.05 &   & \\
KOI-1700 & 14.1 & 2013/07/27 & LP600 & 3.6 & 0.29$\pm$0.06 & 289$\pm$2 & 1.07$\pm$0.26 &   & yes\\ 
KOI-1784 & 13.4 & 2013/07/28 & LP600 & 4.7 & 0.33$\pm$0.06 & 286$\pm$6 & 0.58$\pm$0.13 & K15, W15  & yes\\ 
KOI-1798 & 14.2 & 2013/08/13 & LP600 & 2.4 & 3.81$\pm$0.06 & 186$\pm$2 & 3.75$\pm$0.21 &   & \\ 
KOI-1830 & 14.2 & 2013/07/27 & LP600 & 3.9 & 0.46$\pm$0.06 & 319$\pm$4 & 1.29$\pm$0.17 &   & \\ 
KOI-1950 & 15.7 & 2013/08/22 & LP600 & 3.6 & 3.35$\pm$0.06 & 326$\pm$2 & 1.69$\pm$0.05 &   & \\ 
KOI-1985 & 13.4 & 2013/07/24 & LP600 & 2.2 & 2.82$\pm$0.06 & 156$\pm$2 & 4.19$\pm$0.09 & K16  & \\ 
KOI-1989 & 13.1 & 2013/08/14 & LP600 & 2.9 & 1.12$\pm$0.06 & 41$\pm$2 & 3.49$\pm$0.16 &   & yes\\ 
KOI-2014 & 15.4 & 2013/10/23 & LP600 & 3.1 & 3.75$\pm$0.06 & 267$\pm$2 & 2.50$\pm$0.04 &   & \\ 
KOI-2019 & 15.4 & 2013/08/22 & LP600 & 3.3 & 4.01$\pm$0.06 & 105$\pm$2 & 2.61$\pm$0.21 &   & \\ 
KOI-2055 & 14.3 & 2013/07/25 & LP600 & 2.2 & 3.80$\pm$0.06 & 57$\pm$2 & 4.09$\pm$0.05 &   & \\ 
KOI-2056 & 14.3 & 2013/08/16 & LP600 & 4.9 & 3.87$\pm$0.06 & 131$\pm$2 & 3.37$\pm$0.11 &   & \\ 
KOI-2069 & 13.6 & 2013/08/14 & LP600 & 2.7 & 1.12$\pm$0.06 & 108$\pm$2 & 4.24$\pm$0.51 &   & \\ 
KOI-2083 & 13.4 & 2013/07/28 & LP600 & 2.5 & 0.26$\pm$0.06 & 176$\pm$2 & 1.03$\pm$0.18 &   & yes\\ 
KOI-2115 & 15.8 & 2013/07/21 & LP600 & 2.4 & 3.59$\pm$0.06 & 243$\pm$2 & 2.75$\pm$0.12 &   & \\ 
KOI-2156 & 15.3 & 2013/08/18 & LP600 & 2.0 & 3.35$\pm$0.06 & 303$\pm$2 & 2.64$\pm$0.05 & G16, K16  & \\ 
KOI-2247 & 14.0 & 2013/07/27 & LP600 & 2.3 & 1.90$\pm$0.06 & 355$\pm$2 & 5.12$\pm$0.21 &   & \\ 
KOI-2314 & 14.4 & 2013/07/27 & LP600 & 4.6 & 4.14$\pm$0.06 & 201$\pm$2 & 3.45$\pm$0.22 &   & \\ 
KOI-2317 & 14.1 & 2013/07/27 & LP600 & 4.6 & 1.51$\pm$0.06 & 110$\pm$2 & 4.93$\pm$0.19 &   & yes\\ 
KOI-2377 & 14.5 & 2013/07/24 & LP600 & 4.3 & 4.11$\pm$0.06 & 326$\pm$2 & 4.04$\pm$0.12 &  & yes \\ 
KOI-2380 & 14.0 & 2013/07/24 & LP600 & 4.0 & 4.01$\pm$0.06 & 250$\pm$2 & 2.46$\pm$0.08 &   & \\ 
KOI-2421 & 14.0 & 2013/07/25 & LP600 & 4.1 & 4.07$\pm$0.06 & 132$\pm$2 & 3.87$\pm$0.18 & D14  & \\ 
KOI-2469 & 14.7 & 2013/08/18 & LP600 & 3.8 & 4.18$\pm$0.06 & 114$\pm$2 & 2.44$\pm$0.17 &   & \\ 
KOI-2493 & 15.0 & 2013/08/22 & LP600 & 2.4 & 2.69$\pm$0.06 & 300$\pm$2 & 2.68$\pm$0.04 &   & \\ 
KOI-2516 & 13.1 & 2013/07/27 & LP600 & 4.0 & 3.42$\pm$0.06 & 84$\pm$2 & 5.93$\pm$0.04 & D14  & \\ 
KOI-2542 & 14.8 & 2013/08/21 & LP600 & 3.4 & 0.88$\pm$0.06 & 22$\pm$3 & 1.20$\pm$0.19 & G16, K16  & yes\\ 
KOI-2551 & 15.5 & 2013/10/23 & LP600 & 3.5 & 2.69$\pm$0.06 & 197$\pm$2 & 1.93$\pm$0.03 &   & \\ 
KOI-2601 & 13.8 & 2013/08/13 & LP600 & 3.8 & 1.44$\pm$0.06 & 297$\pm$2 & 3.61$\pm$0.14 &   & \\ 
KOI-2664 & 15.3 & 2013/08/18 & LP600 & 4.4 & 1.17$\pm$0.06 & 90$\pm$2 & 0.83$\pm$0.09 &   & yes\\ 
KOI-2681 & 15.7 & 2013/10/23 & LP600 & 4.2 & 1.10$\pm$0.06 & 161$\pm$2 & 1.25$\pm$0.11 &   & yes\\ 
KOI-2707 & 14.2 & 2013/07/24 & LP600 & 2.2 & 3.28$\pm$0.06 & 217$\pm$2 & 4.71$\pm$0.16 &   & \\ 
 &  &  &  & 2.9 & 3.87$\pm$0.06 & 182$\pm$2 & 3.64$\pm$0.11 &   & \\ 
KOI-2722 & 13.1 & 2013/08/14 & LP600 & 3.7 & 3.27$\pm$0.06 & 282$\pm$2 & 5.88$\pm$0.14 & D14  & yes\\ 
KOI-2743 & 13.5 & 2013/08/14 & LP600 & 3.1 & 2.36$\pm$0.06 & 182$\pm$2 & 3.79$\pm$0.08 &   & \\ 
KOI-2779 & 14.8 & 2013/10/22 & LP600 & 2.9 & 0.98$\pm$0.06 & 61$\pm$2 & 2.54$\pm$0.38 &   & \\ 
KOI-2838 & 13.2 & 2013/08/13 & LP600 & 2.4 & 1.74$\pm$0.06 & 197$\pm$2 & 5.92$\pm$0.35 & D14  & \\ 
KOI-2859 & 13.6 & 2013/08/16 & LP600 & 3.1 & 0.47$\pm$0.06 & 282$\pm$5 & 2.12$\pm$0.23 &   & yes\\ 
KOI-2914 & 12.1 & 2013/07/24 & LP600 & 2.4 & 3.80$\pm$0.06 & 231$\pm$2 & 5.64$\pm$0.06 & D14  & \\ 
KOI-2949 & 13.1 & 2013/07/29 & LP600 & 2.2 & 2.36$\pm$0.06 & 311$\pm$2 & 4.08$\pm$0.32 &   & \\ 
KOI-2971 & 12.6 & 2013/07/24 & LP600 & 3.9 & 0.53$\pm$0.06 & 209$\pm$4 & 1.33$\pm$0.18 &      & yes\\ 
KOI-2984 & 12.9 & 2013/07/24 & LP600 & 4.1 & 3.47$\pm$0.06 & 33$\pm$2 & 4.34$\pm$0.05 & D14  & \\ 
KOI-3029 & 14.7 & 2013/08/18 & LP600 & 3.1 & 0.28$\pm$0.06 & 272$\pm$2 & 0.68$\pm$0.22 & & yes\\
KOI-3041 & 14.0 & 2013/07/25 & LP600 & 3.2 & 2.03$\pm$0.06 & 128$\pm$2 & 4.64$\pm$0.17 &   & \\ 
KOI-3255 & 13.9 & 2013/07/27 & LP600 & 2.2 & 3.15$\pm$0.06 & 44$\pm$2 & 4.87$\pm$0.05 & E15, K16  & \\ 
KOI-3277 & 12.9 & 2013/07/24 & LP600 & 2.1 & 2.45$\pm$0.06 & 355$\pm$2 & 5.79$\pm$0.23 &   & \\ 
 &  &  &  & 3.8 & 3.41$\pm$0.06 & 353$\pm$2 & 5.00$\pm$0.12 &   & \\ 
KOI-3284 & 13.8 & 2013/08/22 & LP600 & 4.2 & 3.94$\pm$0.06 & 4$\pm$2 & 2.42$\pm$0.08 & To15, E15, G16, K16  & \\ 
KOI-3288 & 14.0 & 2013/07/24 & LP600 & 2.4 & 3.17$\pm$0.06 & 75$\pm$2 & 4.32$\pm$0.12 &   & \\ 
 &  &  &  & 2.1 & 3.50$\pm$0.06 & 80$\pm$2 & 4.62$\pm$0.16 &   & \\ 
KOI-3309 & 14.4 & 2013/08/17 & LP600 & 3.2 & 3.71$\pm$0.06 & 42$\pm$2 & 2.78$\pm$0.04 & K16  & \\ 
KOI-3377 & 14.9 & 2013/08/22 & LP600 & 2.7 & 1.45$\pm$0.06 & 58$\pm$2 & 4.26$\pm$0.19 &   & yes\\ 
KOI-3401 & 14.2 & 2013/07/28 & LP600 & 4.3 & 0.65$\pm$0.06 & 94$\pm$3 & 0.89$\pm$0.20 &   & \\ 
KOI-3439 & 14.0 & 2013/07/24 & LP600 & 2.8 & 3.42$\pm$0.06 & 228$\pm$2 & 3.97$\pm$0.05 &   & \\ 
KOI-3444 & 13.0 & 2013/07/25 & LP600 & 2.6 & 1.11$\pm$0.06 & 8$\pm$2 & 3.32$\pm$0.17 & LB14, W15, G16, K16  & \\ 
 &  &  &  & 3.2 & 3.55$\pm$0.06 & 262$\pm$2 & 3.41$\pm$0.23 & LB14, W15, G16, K16  & \\ 
KOI-3459 & 15.0 & 2013/08/21 & LP600 & 3.3 & 3.35$\pm$0.06 & 124$\pm$2 & 2.37$\pm$0.04 &   & \\ 
KOI-3460 & 14.3 & 2013/07/27 & LP600 & 2.5 & 1.24$\pm$0.06 & 153$\pm$2 & 5.08$\pm$0.13 &   & \\ 
 &  &  &  & 2.8 & 2.47$\pm$0.06 & 231$\pm$2 & 5.52$\pm$0.11 &   & \\ 
KOI-3486 & 14.1 & 2013/07/24 & LP600 & 2.7 & 4.16$\pm$0.06 & 260$\pm$2 & 4.06$\pm$0.12 &   & \\ 
KOI-3500 & 13.0 & 2013/07/25 & LP600 & 4.1 & 2.54$\pm$0.06 & 137$\pm$2 & 4.01$\pm$0.05 &   & \\ 
KOI-3946 & 13.1 & 2013/08/13 & LP600 & 3.9 & 4.27$\pm$0.06 & 61$\pm$2 & 5.26$\pm$0.13 &   & \\ 
KOI-4053 & 12.6 & 2013/07/25 & LP600 & 3.2 & 4.11$\pm$0.06 & 302$\pm$2 & 5.51$\pm$0.12 &   & \\ 
KOI-4098 & 13.5 & 2013/08/15 & LP600 & 4.5 & 0.78$\pm$0.06 & 174$\pm$3 & 1.10$\pm$0.22 &   & \\ 
KOI-4166 & 15.0 & 2013/08/18 & LP600 & 2.0 & 3.54$\pm$0.06 & 157$\pm$2 & 3.29$\pm$0.04 &   & \\ 
KOI-4194 & 15.0 & 2013/08/21 & LP600 & 2.3 & 2.17$\pm$0.06 & 290$\pm$2 & 3.41$\pm$0.07 &   & \\ 
KOI-4208 & 15.3 & 2013/08/17 & LP600 & 3.3 & 0.99$\pm$0.06 & 234$\pm$2 & 2.57$\pm$0.29 &   & \\ 
KOI-4226 & 12.6 & 2013/07/24 & LP600 & 4.7 & 2.49$\pm$0.06 & 267$\pm$2 & 4.18$\pm$0.03 & K16  & \\ 
KOI-4274 & 15.1 & 2013/08/22 & LP600 & 3.1 & 3.26$\pm$0.06 & 207$\pm$2 & 3.71$\pm$0.11 &   & \\ 
 &  &  &  & 2.4 & 4.54$\pm$0.06 & 327$\pm$2 & 4.11$\pm$0.13 &   & \\ 
KOI-4313 & 14.1 & 2013/08/13 & LP600 & 2.6 & 2.88$\pm$0.06 & 81$\pm$2 & 4.19$\pm$0.05 &   & \\
KOI-4409 & 12.3 & 2013/07/24 & LP600 & 2.3 & 2.89$\pm$0.06 & 139$\pm$2 & 6.10$\pm$0.10 &   & \\ 
KOI-4443 & 13.7 & 2013/08/13 & LP600 & 2.2 & 3.41$\pm$0.06 & 26$\pm$2 & 5.00$\pm$0.13 &   & \\ 
KOI-4495 & 15.2 & 2013/08/22 & LP600 & 3.9 & 3.04$\pm$0.06 & 58$\pm$2 & 4.73$\pm$0.09 &   & \\ 
KOI-4523 & 14.6 & 2013/08/21 & LP600 & 2.4 & 3.94$\pm$0.06 & 100$\pm$2 & 2.61$\pm$0.33 &   & \\ 
KOI-4699 & 12.8 & 2013/08/13 & LP600 & 2.8 & 4.01$\pm$0.06 & 285$\pm$2 & 5.93$\pm$0.21 &   & \\ 
KOI-4768 & 15.4 & 2013/10/23 & LP600 & 2.3 & 1.30$\pm$0.06 & 159$\pm$2 & 3.99$\pm$0.26 &   & yes\\ 
KOI-4797 & 15.3 & 2013/08/22 & LP600 & 3.5 & 3.93$\pm$0.06 & 77$\pm$2 & 3.37$\pm$0.11 &  & \\ 
KOI-4812 & 15.5 & 2013/10/23 & LP600 & 3.6 & 3.15$\pm$0.06 & 100$\pm$2 & 1.84$\pm$0.03 &   & \\
KOI-4813 & 13.3 & 2013/07/28 & LP600 & 4.6 & 4.03$\pm$0.06 & 146$\pm$2 & 3.34$\pm$0.09 &  & \\ 

\enddata
\tablenotetext{}{\textbf{Notes:} References for previous detections are denoted with the following codes: \citealt{A12} (A12); \citealt{D14} (D14); \citealt{E15} (E15); \citealt{G16} (G16); \citealt{K15} (K15); \citealt{K16} (K16); \citealt{LB12} (LB12); \citealt{LB14} (LB14); \citealt{To15} (To15); \citealt{W15} (W15).}
\end{deluxetable*}

\clearpage

\begin{table}
\begin{center}
\caption{\label{tab:table_keck_nirc2}Full Keck-NIRC2 Observation List and Detected Companions}
\begin{tabular}{llccrrl}
\tableline
\tableline
KOI        & ObsID       & Filter & Separation    & P.A.      & Mag. Diff.    & Notes      \\
           &             &        &  (arcsec)     & (deg.)    & (mag)         &            \\
\tableline
KOI-190    & 2015 Jul 25 & Kp     & 0.180$\pm$0.010 & 109$\pm$3 & 0.64$\pm$0.14 & Table \ref{tab:comp_low} \\
KOI-425    & 2015 Jul 25 & Kp     & 0.490$\pm$0.010 & 343$\pm$1 & 0.83$\pm$0.06 & Table \ref{tab:comp_low} \\
KOI-628    & 2013 Aug 25 & Kp     & 1.828$\pm$0.005 & 311$\pm$1 & 3.87$\pm$0.06 & PI        \\
           &             &        & 2.752$\pm$0.005 & 239$\pm$1 & 3.00$\pm$0.06 & c          \\
KOI-931    & 2015 Jul 25 & Kp     & 1.261$\pm$0.005 & 177$\pm$1 & 3.23$\pm$0.06 & Table \ref{tab:comp_low} \\
KOI-987    & 2013 Aug 25 & Kp     & 1.977$\pm$0.005 & 226$\pm$1 & 2.24$\pm$0.06 & PI        \\
KOI-1066   & 2014 Aug 17 & Kp     & 1.687$\pm$0.005 & 231$\pm$1 & 2.95$\pm$0.07 & Table \ref{tab:comp_low}           \\
KOI-1067   & 2014 Aug 17 & Kp     & 2.923$\pm$0.005 & 142$\pm$1 & 3.79$\pm$0.11 & Table \ref{tab:comp_low}           \\
KOI-1112   & 2013 Aug 24 & Kp     & 3.063$\pm$0.005 & 172$\pm$1 & 2.76$\pm$0.07 & Table \ref{tab:comp_low}           \\
KOI-1151   & 2014 Aug 17 & Kp     & 0.756$\pm$0.010 & 307$\pm$1 & 2.41$\pm$0.06 & PI        \\
KOI-1214   & 2014 Aug 17 & Kp     & 0.347$\pm$0.010 & 136$\pm$2 & 2.46$\pm$0.06 & Table \ref{tab:comp_low}           \\
KOI-1359   & 2013 Jun 25 & Ks     & 1.387$\pm$0.005 & 332$\pm$1 & 2.17$\pm$0.06 & PI        \\
KOI-1375   & 2014 Aug 17 & Kp     & 0.785$\pm$0.010 &  27$\pm$1 & 3.39$\pm$0.07 & PI        \\
KOI-1442   & 2013 Aug 25 & Kp     & 2.113$\pm$0.005 &  71$\pm$1 & 3.63$\pm$0.06 & PI        \\
KOI-1546   & 2013 Aug 24 & Kp     & 0.600$\pm$0.010 &  90$\pm$1 & 0.73$\pm$0.05 & Table \ref{tab:comp_high}           \\
           &             &        & 2.918$\pm$0.005 &   4$\pm$1 & 2.95$\pm$0.08 & Table \ref{tab:comp_low}           \\
           &             &        & 4.108$\pm$0.005 & 165$\pm$1 & 3.48$\pm$0.06 & Table \ref{tab:comp_low}        \\
KOI-1700   & 2014 Aug 17 & Kp     & 0.274$\pm$0.010 & 288$\pm$2 & 0.55$\pm$0.06 & Table \ref{tab:comp_low}           \\
KOI-1784   & 2015 Jul 25 & Kp     & 0.279$\pm$0.010 & 291$\pm$2 & 0.78$\pm$0.06 & Table \ref{tab:comp_low}           \\
KOI-1845   & 2013 Jun 25 & Ks     & 1.996$\pm$0.005 &  79$\pm$1 & 2.89$\pm$0.06 & PI        \\
           &             &        & 2.963$\pm$0.005 & 348$\pm$1 & 4.40$\pm$0.09 & c          \\
KOI-1884   & 2013 Aug 24 & K      & 0.934$\pm$0.010 &  95$\pm$1 & 2.31$\pm$0.06 & PI        \\
           &             &        & 1.838$\pm$0.005 &  82$\pm$1 & 2.73$\pm$0.06 & b          \\
           &             &        & 2.567$\pm$0.005 & 327$\pm$1 & 3.20$\pm$0.14 & b          \\
KOI-1891   & 2013 Aug 24 & Kp     & 2.069$\pm$0.005 & 211$\pm$1 & 4.60$\pm$0.07 & PI        \\
KOI-1989   & 2015 Jul 25 & Kp     & 0.817$\pm$0.010 &  40$\pm$1 & 2.92$\pm$0.06 & Table \ref{tab:comp_low}           \\
KOI-2009   & 2013 Aug 25 & Kp     & 1.511$\pm$0.005 & 178$\pm$1 & 2.75$\pm$0.06 & PI        \\
KOI-2083   & 2014 Aug 17 & Kp     & 0.252$\pm$0.010 & 166$\pm$1 & 1.60$\pm$0.05 & Table \ref{tab:comp_low}           \\
KOI-2159   & 2014 Aug 17 & Kp     & 2.018$\pm$0.005 & 324$\pm$1 & 2.48$\pm$0.06 & PI        \\
KOI-2317   & 2015 Jul 25 & Kp     & 1.512$\pm$0.005 & 112$\pm$1 & 3.92$\pm$0.06 & Table \ref{tab:comp_low}           \\
KOI-2363   & 2014 Aug 17 & Kp     & 1.952$\pm$0.005 & 357$\pm$1 & 5.04$\pm$0.09 & b          \\
KOI-2377   & 2013 Aug 24 & Kp     & 2.186$\pm$0.005 & 335$\pm$1 & 0.63$\pm$0.07 & Table \ref{tab:comp_high}           \\
           &             &        & 2.544$\pm$0.005 &  42$\pm$1 & 3.75$\pm$0.12 & b          \\
           &             &        & 3.911$\pm$0.005 & 316$\pm$1 & 3.55$\pm$0.17 & Table \ref{tab:comp_low}           \\
KOI-2413   & 2013 Aug 25 & Kp     & 0.308$\pm$0.010 & 250$\pm$2 & 0.17$\pm$0.06 & PI        \\
KOI-2443   & 2013 Jun 25 & Ks     & 1.383$\pm$0.005 & 164$\pm$1 & 3.63$\pm$0.06 & PI        \\
KOI-2542   & 2015 Jul 25 & Kp     & 0.765$\pm$0.010 &  29$\pm$1 & 0.60$\pm$0.05 & Table \ref{tab:comp_low}           \\
KOI-2664   & 2015 Jul 25 & Kp     & 1.180$\pm$0.005 &  90$\pm$1 & 1.10$\pm$0.06 & Table \ref{tab:comp_low}           \\
KOI-2681   & 2014 Aug 17 & Kp     & 1.132$\pm$0.005 & 148$\pm$1 & 0.43$\pm$0.06 & Table \ref{tab:comp_low}           \\
KOI-2705   & 2013 Aug 24 & Kp     & 1.900$\pm$0.005 & 304$\pm$1 & 2.58$\pm$0.07 & Table \ref{tab:comp_high}           \\
KOI-2711   & 2014 Aug 17 & Kp     & 0.466$\pm$0.010 & 149$\pm$1 & 0.12$\pm$0.06 & Table \ref{tab:comp_high}           \\
KOI-2722   & 2014 Aug 17 & Kp     & 3.226$\pm$0.005 & 283$\pm$1 & 3.77$\pm$0.06 & Table \ref{tab:comp_low}           \\
KOI-2837   & 2013 Aug 24 & Kp     & 0.347$\pm$0.010 & 138$\pm$2 & 0.20$\pm$0.06 & Table \ref{tab:comp_high}           \\
KOI-2859   & 2014 Aug 17 & Kp     & 0.454$\pm$0.010 & 291$\pm$1 & 2.89$\pm$0.06 & Table \ref{tab:comp_low}           \\
KOI-2904   & 2013 Aug 24 & Kp     & 0.700$\pm$0.010 & 226$\pm$1 & 2.45$\pm$0.06 & Table \ref{tab:comp_high}           \\
KOI-2971   & 2015 Jul 25 & Kp     & 0.301$\pm$0.010 & 274$\pm$2 & 3.57$\pm$0.06 & Table \ref{tab:comp_low}          \\
           &             &        & 3.564$\pm$0.005 &  38$\pm$1 & 5.93$\pm$0.17 & b           \\
KOI-3029   & 2014 Aug 17 & Kp     & 0.251$\pm$0.010 & 264$\pm$2 & 0.06$\pm$0.06 & Table \ref{tab:comp_low}           \\
           &             &        & 1.734$\pm$0.005 & 356$\pm$1 & 4.49$\pm$0.07 & b          \\
           &             &        & 2.552$\pm$0.005 &   4$\pm$1 & 3.44$\pm$0.07 & b          \\
KOI-3069   & 2013 Aug 24 & Kp     & 1.785$\pm$0.005 & 108$\pm$1 & 1.26$\pm$0.06 & Table \ref{tab:comp_high}           \\
KOI-3377   & 2015 Jul 25 & Kp     & 0.264$\pm$0.010 & 335$\pm$2 & 0.49$\pm$0.06 & b          \\
           &             &        & 1.405$\pm$0.005 &  50$\pm$1 & 3.74$\pm$0.06 & Table \ref{tab:comp_low}           \\
KOI-4004   & 2015 Jul 25 & Kp     & 1.958$\pm$0.005 & 218$\pm$1 & 2.37$\pm$0.08 & Table \ref{tab:comp_high}           \\
KOI-4209   & 2013 Aug 24 & Kp     & 0.975$\pm$0.010 & 205$\pm$1 & 0.57$\pm$0.06 & Table \ref{tab:comp_high}           \\
KOI-4292   & 2013 Aug 25 & Kp     & 1.951$\pm$0.005 &  30$\pm$1 & 4.54$\pm$0.08 & b          \\
KOI-4331   & 2014 Aug 17 & Kp     & 0.347$\pm$0.010 & 102$\pm$2 & 0.13$\pm$0.05 & Table \ref{tab:comp_high}           \\
KOI-4407   & 2013 Aug 25 & Kp     & 2.455$\pm$0.005 & 300$\pm$1 & 1.89$\pm$0.06 & Table \ref{tab:comp_high}           \\
           &             &        & 2.662$\pm$0.005 & 311$\pm$1 & 4.65$\pm$0.34 & b          \\
KOI-4463   & 2013 Aug 24 & Kp     & 2.462$\pm$0.005 & 144$\pm$1 & -0.26$\pm$0.07 & Table \ref{tab:comp_high}           \\
KOI-4634   & 2015 Jul 25 & Kp     & 0.282$\pm$0.010 & 276$\pm$2 & 0.65$\pm$0.06 & Table \ref{tab:comp_high}           \\
KOI-4768   & 2014 Aug 17 & Kp     & 1.326$\pm$0.005 & 160$\pm$1 & 2.61$\pm$0.07 & Table \ref{tab:comp_low}           \\
KOI-4871   & 2014 Aug 17 & Kp     & 0.921$\pm$0.010 & 334$\pm$1 & 3.04$\pm$0.06 & Table \ref{tab:comp_high}           \\
\tableline
\end{tabular}
\tablenotetext{1}{PI = Confirmation of $<5\sigma$ significance companions in Paper I. }
\tablenotetext{2}{Companion not detected in Robo-AO image.}
\tablenotetext{3}{Additional detection not reported in Paper I. (Angular separation exceeds 2\farcs5.)}
\end{center}
\end{table}



\clearpage

\begin{deluxetable*}{lllcc}
\tablewidth{0pt}
\tablecaption{Full Robo-AO Observation List}
\tablehead{
\colhead{KOI}           & \colhead{$m_{i}$/mags}      &
\colhead{ObsID}          &
\colhead{Obs. qual.$^*$}          & \colhead{Companion?}}
\startdata
K00004.01 & 11.3 & 2012/07/16 & high ($i'$-band) & yes \\ 
K00042.01 & 9.2 & 2013/07/28  & high & yes \\ 
K00116.01 & 12.7 & 2013/07/24  & high &  \\ 
K00123.01 & 12.2 & 2013/07/25  & high &  \\ 
K00127.01 & 13.7 & 2013/08/15  & high &  \\ 
K00138.01 & 13.9 & 2013/07/29  & medium &  \\ 
K00150.01 & 13.6 & 2013/07/27  & high &  \\ 
K00151.01 & 13.8 & 2013/07/27  & high & yes \\ 
K00155.01 & 13.3 & 2013/08/13  & high & yes \\ 
K00183.01 & 14.1 & 2013/08/15  & medium &  \\ 
K00188.01 & 14.5 & 2013/08/22  & medium &  \\ 
K00189.01 & 14.1 & 2013/08/17  & medium &  \\ 
K00190.01 & 13.9 & 2013/07/27  & medium & yes \\ 
K00192.01 & 14.1 & 2013/08/15  & medium &  \\ 
K00202.01 & 14.1 & 2013/08/13  & low &  \\ 
K00205.01 & 14.2 & 2013/07/27  & high &  \\ 
K00206.01 & 14.3 & 2013/07/25  & medium &  \\ 
K00208.01 & 14.8 & 2013/08/21  & low &  \\ 
K00221.01 & 14.4 & 2013/07/27  & medium &  \\ 
K00227.01 & 13.7 & 2013/07/27  & low & yes \\ 
K00234.01 & 14.1 & 2013/08/15  & medium &  \\ 
K00235.01 & 14.1 & 2013/07/29  & medium &  \\ 
K00239.01 & 14.6 & 2013/08/22  & medium &  \\ 
K00245.01 &  & 2013/07/28  & high &  \\ 
K00251.01 & 14.1 & 2013/08/21  & medium & yes \\ 
K00257.01 & 10.7 & 2013/07/27  & high &  \\ 
K00258.01 & 9.8 & 2012/07/18 & high ($i'$-band) & yes \\ 
K00262.01 & 10.3 & 2012/07/28 & high ($i'$-band) &  \\ 
K00265.01 & 11.9 & 2013/08/13  & high &  \\ 
K00271.01 & 11.4 & 2012/08/02 & high ($i'$-band) &  \\ 
K00274.01 & 11.3 & 2013/07/28  & high &  \\ 
K00284.01 & 11.7 & 2013/07/27  & medium & yes \\ 
K00285.01 &  & 2013/08/13  & medium & yes \\ 
K00289.01 & 12.5 & 2013/08/13  & high &  \\ 
K00292.01 & 12.7 & 2013/08/18  & low &  \\ 
K00295.01 & 12.2 & 2013/08/17  & high &  \\ 
K00298.01 & 12.4 & 2013/08/15  & high & yes \\ 
K00304.01 & 12.4 & 2013/07/29  & high &  \\ 
K00318.01 & 12.1 & 2013/08/15  & high &  \\ 
K00326.01 & 13.0 & 2013/07/29  & high &  \\ 
K00338.01 & 13.1 & 2013/07/28  & high &  \\ 
K00346.01 & 13.2 & 2013/07/27  & high &  \\ 
K00351.01 & 13.7 & 2013/08/15  & medium &  \\ 
K00354.01 & 13.1 & 2013/08/15  & high &  \\ 
K00355.01 & 13.0 & 2013/08/15  & high &  \\ 
K00367.01 & 10.9 & 2013/07/27  & high &  \\ 
K00369.01 & 11.9 & 2013/07/24  & high &  \\ 
K00370.01 & 11.8 & 2013/08/14  & medium &  \\ 
K00374.01 & 12.0 & 2013/08/15  & high &  \\ 
K00375.01 & 13.1 & 2013/08/17  & high &  \\ 
K00379.01 & 13.2 & 2013/07/27  & high & yes \\ 
K00387.01 & 13.2 & 2013/07/29  & medium & yes \\ 
K00410.01 & 14.3 & 2013/07/29  & medium &  \\ 
K00412.01 & 14.1 & 2013/07/25  & low &  \\ 
K00417.01 & 14.6 & 2013/10/22  & low &  \\ 
K00419.01 & 14.3 & 2013/08/13  & medium &  \\ 
K00420.01 & 13.9 & 2013/08/13  & high &  \\ 
K00421.01 & 14.7 & 2013/08/21  & low &  \\ 
K00425.01 & 14.5 & 2013/08/22  & low & yes \\ 
K00429.01 & 14.2 & 2013/07/28  & medium &  \\ 
K00432.01 & 14.1 & 2013/08/17  & medium &  \\ 
K00433.01 & 14.6 & 2013/08/21  & low &  \\ 
K00438.01 & 13.8 & 2013/08/18  & medium & yes \\ 
K00443.01 & 14.0 & 2013/07/24  & medium &  \\ 
K00446.01 & 14.0 & 2013/07/27  & medium &  \\ 
K00448.01 & 14.5 & 2013/08/22  & medium &  \\ 
K00460.01 & 14.5 & 2013/10/23  & medium &  \\ 
K00470.01 & 14.5 & 2013/10/23  & medium &  \\ 
K00473.01 & 14.4 & 2013/10/22  & low &  \\ 
K00475.01 & 14.5 & 2012/09/01  & medium &  \\ 
K00477.01 & 14.4 & 2013/08/22  & low &  \\ 
K00480.01 & 14.1 & 2013/08/17  & medium &  \\ 
K00484.01 & 14.2 & 2013/08/17  & medium &  \\ 
K00487.01 & 14.3 & 2013/08/18  & medium &  \\ 
K00488.01 & 14.5 & 2013/08/18  & medium &  \\ 
K00492.01 & 14.2 & 2013/07/24  & high &  \\ 
K00494.01 & 14.6 & 2013/08/22  & medium &  \\ 
K00496.01 & 14.1 & 2013/08/13  & medium &  \\ 
K00499.01 & 14.0 & 2013/07/25  & low &  \\ 
K00500.01 & 14.4 & 2013/08/21  & medium &  \\ 
K00501.01 & 14.4 & 2013/07/24  & medium &  \\ 
K00503.01 & 14.5 & 2013/08/22  & low &  \\ 
K00504.01 & 14.3 & 2013/07/25  & low &  \\ 
K00505.01 & 13.9 & 2013/07/29  & medium &  \\ 
K00507.01 & 14.6 & 2013/08/22  & medium & yes \\ 
K00512.01 & 14.6 & 2013/08/22  & low &  \\ 
K00518.01 & 14.0 & 2013/08/15  & low &  \\ 
K00521.01 & 14.5 & 2013/08/14  & medium & yes \\ 
K00525.01 & 14.3 & 2013/07/24  & medium &  \\ 
K00526.01 & 14.2 & 2013/08/14  & medium &  \\ 
K00535.01 & 14.2 & 2013/08/13  & medium &  \\ 
K00536.01 & 14.3 & 2013/08/15  & medium &  \\ 
K00538.01 & 14.4 & 2013/08/15  & medium &  \\ 
K00554.01 & 14.4 & 2013/08/13  & medium &  \\ 
K00558.01 & 14.6 & 2013/08/22  & low & yes \\ 
K00563.01 & 14.3 & 2013/07/27  & high &  \\ 
K00577.01 & 14.1 & 2013/07/29  & medium &  \\ 
K00584.01 & 13.9 & 2012/08/05  & medium & yes \\ 
K00586.01 & 14.4 & 2013/07/29  & medium &  \\ 
K00587.01 & 14.3 & 2013/07/27  & high &  \\ 
K00588.01 & 14.0 & 2013/07/28  & low &  \\ 
K00592.01 & 14.1 & 2013/08/14  & high & yes \\ 
K00607.01 & 14.2 & 2013/08/13  & medium &  \\ 
K00610.01 & 14.1 & 2013/08/21  & medium &  \\ 
K00614.01 & 14.3 & 2013/07/29  & low & yes \\ 
K00617.01 & 14.4 & 2013/07/28  & low &  \\ 
K00618.01 & 14.7 & 2013/10/22  & low &  \\ 
K00641.01 & 13.1 & 2013/07/24  & high & yes \\ 
K00645.01 & 13.5 & 2013/07/25  & medium & yes \\ 
K00652.01 & 13.3 & 2013/07/28  & high & yes \\ 
K00670.01 & 13.6 & 2013/08/13  & high &  \\ 
K00672.01 & 13.8 & 2013/08/14  & medium &  \\ 
K00678.01 & 13.0 & 2013/07/29  & medium &  \\ 
K00683.01 & 13.5 & 2013/07/25  & medium &  \\ 
K00693.01 & 13.8 & 2013/08/13  & low &  \\ 
K00697.01 & 13.5 & 2013/08/15  & high & yes \\ 
K00728.01 & 15.2 & 2013/10/23  & low &  \\ 
K00730.01 & 15.1 & 2013/08/21  & low & yes \\ 
K00735.01 & 15.3 & 2013/10/23  & low &  \\ 
K00751.01 & 15.6 & 2013/08/22  & low &  \\ 
K00753.01 & 15.2 & 2013/10/22  & low &  \\ 
K00758.01 & 15.0 & 2013/10/22  & low &  \\ 
K00762.01 & 15.2 & 2013/08/21  & low &  \\ 
K00767.01 & 14.8 & 2013/08/22  & low &  \\ 
K00785.01 & 15.3 & 2013/10/23  & low &  \\ 
K00790.01 & 15.1 & 2013/10/23  & low &  \\ 
K00794.01 & 14.8 & 2013/08/21  & low &  \\ 
K00795.01 & 15.4 & 2013/08/22  & low &  \\ 
K00797.01 & 15.5 & 2013/08/22  & low &  \\ 
K00801.01 & 14.8 & 2013/10/25  & medium & yes \\ 
K00813.01 & 15.5 & 2013/08/22  & low & yes \\ 
K00821.01 & 15.3 & 2013/10/23  & low &  \\ 
K00822.01 & 15.6 & 2013/08/22  & low &  \\ 
K00823.01 & 15.0 & 2013/10/23  & low &  \\ 
K00824.01 & 16.1 & 2013/08/21  & low &  \\ 
K00825.01 & 14.9 & 2013/08/22  & low &  \\ 
K00844.01 & 15.3 & 2013/08/22  & low &  \\ 
K00846.01 & 15.3 & 2013/08/22  & low &  \\ 
K00851.01 & 15.1 & 2013/08/17  & low &  \\ 
K00875.01 & 15.2 & 2013/08/22  & low &  \\ 
K00878.01 & 15.0 & 2013/08/22  & low &  \\ 
K00883.01 & 15.4 & 2013/10/23  & low &  \\ 
K00895.01 & 15.1 & 2013/10/23  & low &  \\ 
K00897.01 & 15.0 & 2013/08/18  & low &  \\ 
K00908.01 & 14.8 & 2013/08/21  & low &  \\ 
K00922.01 & 15.1 & 2013/08/21  & low &  \\ 
K00931.01 & 15.1 & 2013/10/23  & low & yes \\ 
K00940.01 & 14.2 & 2013/07/27  & medium &  \\ 
K00941.01 & 15.1 & 2013/10/22  & low &  \\ 
K00945.01 & 14.9 & 2013/08/22  & low &  \\ 
K00969.01 &  & 2013/07/28  & high &  \\ 
K00972.01 & 9.4 & 2013/07/21  & low &  \\ 
K00974.01 & 9.5 & 2013/07/29  & high &  \\ 
K00976.01 &  & 2012/08/03 & high ($i'$-band) & yes \\ 
K00981.01 & 10.5 & 2013/07/28  & high &  \\ 
K00992.01 & 15.0 & 2013/08/18  & low &  \\ 
K00993.01 & 14.0 & 2013/07/25  & medium &  \\ 
K00994.01 & 14.4 & 2013/07/25  & medium &  \\ 
K00998.01 & 15.5 & 2013/08/18  & low &  \\ 
K00999.01 & 15.1 & 2013/08/18  & low & yes \\ 
K01007.01 & 15.0 & 2013/08/18  & low &  \\ 
K01014.01 & 15.4 & 2013/08/21  & low &  \\ 
K01022.01 & 15.5 & 2013/08/18  & low &  \\ 
K01024.01 & 14.0 & 2013/07/27  & high &  \\ 
K01029.01 & 14.5 & 2013/08/18  & medium &  \\ 
K01030.01 & 15.2 & 2013/08/21  & low &  \\ 
K01031.01 & 14.9 & 2013/08/18  & low &  \\ 
K01059.01 & 14.6 & 2013/08/22  & medium &  \\ 
K01061.01 & 14.3 & 2013/08/20  & low & yes \\ 
K01066.01 & 15.4 & 2013/08/21  & low & yes \\ 
K01067.01 & 14.5 & 2013/10/23  & medium & yes \\ 
K01083.01 & 15.1 & 2013/08/22  & low &  \\ 
K01086.01 & 14.4 & 2013/08/14  & low &  \\ 
K01094.01 & 15.5 & 2013/08/18  & low &  \\ 
K01099.01 & 15.2 & 2013/08/22  & low &  \\ 
K01101.01 & 15.4 & 2013/08/21  & low &  \\ 
K01103.01 & 14.8 & 2013/08/18  & low &  \\ 
K01106.01 & 14.6 & 2013/08/22  & medium &  \\ 
K01108.01 & 14.4 & 2013/07/24  & medium &  \\ 
K01109.01 & 14.5 & 2013/08/22  & low &  \\ 
K01110.01 & 14.6 & 2013/08/21  & low &  \\ 
K01112.01 & 14.4 & 2013/07/27  & medium & yes \\ 
K01117.01 & 12.7 & 2013/07/24  & medium &  \\ 
K01159.01 & 15.0 & 2013/08/22  & low &  \\ 
K01174.01 & 13.0 & 2013/07/29  & medium &  \\ 
K01187.01 & 14.3 & 2013/07/29  & medium &  \\ 
K01192.01 & 14.0 & 2013/07/29  & low &  \\ 
K01199.01 & 14.5 & 2013/10/23  & low &  \\ 
K01205.01 & 14.3 & 2013/07/24  & medium &  \\ 
K01206.01 & 13.4 & 2013/07/28  & medium &  \\ 
K01209.01 & 14.9 & 2013/08/21  & low &  \\ 
K01210.01 & 14.2 & 2013/07/28  & medium &  \\ 
K01214.01 & 14.4 & 2013/07/24  & medium & yes \\ 
K01219.01 & 14.2 & 2013/07/29  & low &  \\ 
K01226.01 & 15.0 & 2013/10/23  & low &  \\ 
K01238.01 & 14.3 & 2013/07/25  & medium &  \\ 
K01245.01 & 14.1 & 2013/08/13  & medium &  \\ 
K01281.01 & 14.2 & 2013/08/15  & low &  \\ 
K01300.01 & 13.9 & 2013/08/16  & medium & yes \\ 
K01302.01 & 14.6 & 2013/08/21  & medium &  \\ 
K01304.01 & 15.6 & 2013/10/23  & low &  \\ 
K01310.01 & 14.4 & 2013/08/15  & medium &  \\ 
K01311.01 & 13.3 & 2013/08/14  & medium &  \\ 
K01323.01 & 15.3 & 2013/08/18  & low &  \\ 
K01325.01 & 14.9 & 2013/08/21  & low &  \\ 
K01328.01 & 15.4 & 2013/08/21  & low &  \\ 
K01329.01 & 14.8 & 2013/08/22  & low &  \\ 
K01331.01 & 14.9 & 2013/10/23  & low &  \\ 
K01339.01 & 14.6 & 2013/08/18  & medium &  \\ 
K01341.01 & 14.8 & 2013/08/22  & medium &  \\ 
K01355.01 & 15.7 & 2013/08/22  & low &  \\ 
K01357.01 & 15.3 & 2013/08/18  & low & yes \\ 
K01362.01 & 15.1 & 2013/08/21  & low &  \\ 
K01370.01 & 14.7 & 2013/08/22  & low &  \\ 
K01372.01 & 15.2 & 2013/10/23  & low &  \\ 
K01391.01 & 14.2 & 2013/08/13  & medium &  \\ 
K01397.01 & 14.8 & 2013/08/21  & low & yes \\ 
K01406.01 & 14.5 & 2013/08/15  & low &  \\ 
K01413.01 & 14.2 & 2013/08/14  & medium &  \\ 
K01425.01 & 15.1 & 2013/10/23  & low &  \\ 
K01428.01 & 14.3 & 2013/08/17  & medium &  \\ 
K01431.01 & 13.2 & 2013/08/17  & medium &  \\ 
K01445.01 & 12.2 & 2013/08/14  & high &  \\ 
K01458.01 & 15.5 & 2013/10/22  & low &  \\ 
K01465.01 & 14.1 & 2013/08/17  & medium &  \\ 
K01470.01 & 15.5 & 2013/10/22  & low &  \\ 
K01474.01 & 12.9 & 2013/08/16  & high &  \\ 
K01475.01 & 15.4 & 2013/08/22  & low &  \\ 
K01491.01 & 15.0 & 2013/08/22  & low &  \\ 
K01495.01 & 15.2 & 2013/08/22  & low & yes \\ 
K01499.01 & 14.2 & 2013/07/25  & medium &  \\ 
K01516.01 & 14.7 & 2013/10/23  & low &  \\ 
K01518.01 & 15.0 & 2013/08/22  & low &  \\ 
K01520.01 & 14.3 & 2013/08/16  & medium &  \\ 
K01522.01 & 14.0 & 2013/08/15  & high &  \\ 
K01527.01 & 14.6 & 2013/10/23  & medium &  \\ 
K01531.01 & 12.9 & 2013/08/16  & medium & yes \\ 
K01532.01 & 12.7 & 2013/08/17  & medium &  \\ 
K01533.01 & 13.8 & 2013/08/13  & low &  \\ 
K01534.01 & 13.3 & 2013/07/29  & medium &  \\ 
K01546.01 & 14.2 & 2013/08/19  & low & yes \\ 
K01549.01 & 14.9 & 2013/08/21  & low &  \\ 
K01552.01 & 15.5 & 2013/08/21  & low &  \\ 
K01553.01 & 15.0 & 2013/08/18  & low &  \\ 
K01573.01 & 14.2 & 2013/07/25  & medium & yes \\ 
K01574.01 & 14.4 & 2013/07/28  & low &  \\ 
K01585.01 & 15.2 & 2013/08/22  & medium &  \\ 
K01586.01 & 14.6 & 2013/10/23  & low &  \\ 
K01591.01 & 15.1 & 2013/08/22  & low &  \\ 
K01599.01 & 14.6 & 2013/08/18  & low & yes \\ 
K01601.01 & 14.4 & 2013/07/29  & low &  \\ 
K01602.01 & 14.7 & 2013/08/21  & low &  \\ 
K01603.01 & 14.3 & 2013/07/25  & medium &  \\ 
K01626.01 & 15.1 & 2013/08/21  & low &  \\ 
K01637.01 & 15.8 & 2013/08/21  & low &  \\ 
K01639.01 & 13.4 & 2013/07/28  & high &  \\ 
K01641.01 & 14.8 & 2013/08/21  & low &  \\ 
K01643.01 & 14.5 & 2013/08/22  & medium &  \\ 
K01646.01 & 14.0 & 2013/07/28  & high &  \\ 
K01648.01 & 14.1 & 2013/08/14  & medium &  \\ 
K01683.01 & 14.4 & 2013/08/13  & medium &  \\ 
K01686.01 & 15.1 & 2013/08/18  & low &  \\ 
K01688.01 & 14.3 & 2013/07/25  & medium &  \\ 
K01700.01 & 14.1 & 2013/07/27  & medium & yes \\ 
K01702.01 & 14.9 & 2013/08/22  & low &  \\ 
K01704.01 & 14.6 & 2013/08/21  & low &  \\ 
K01707.01 & 15.1 & 2013/08/22  & low &  \\ 
K01717.01 & 14.3 & 2013/10/25  & low & yes \\ 
K01731.01 & 15.3 & 2013/08/17  & low &  \\ 
K01733.01 & 15.5 & 2013/08/19  & low &  \\ 
K01746.01 & 14.7 & 2013/08/21  & low &  \\ 
K01760.01 & 15.1 & 2013/08/22  & low &  \\ 
K01762.01 & 14.8 & 2013/08/22  & low &  \\ 
K01784.01 & 13.4 & 2013/07/28  & high & yes \\ 
K01786.01 & 14.2 & 2013/08/18  & medium &  \\ 
K01788.01 & 14.1 & 2013/07/25  & medium &  \\ 
K01793.01 & 15.1 & 2013/08/18  & low &  \\ 
K01797.01 & 12.6 & 2013/08/14  & high &  \\ 
K01798.01 & 14.2 & 2013/08/13  & medium & yes \\ 
K01800.01 & 12.2 & 2013/08/15  & medium &  \\ 
K01801.01 & 14.4 & 2013/08/13  & medium &  \\ 
K01808.01 & 12.3 & 2013/07/29  & high &  \\ 
K01809.01 & 12.5 & 2013/08/13  & medium &  \\ 
K01815.01 & 13.5 & 2013/07/24  & high &  \\ 
K01826.01 & 13.2 & 2013/07/25  & medium &  \\ 
K01828.01 & 14.2 & 2013/08/13  & medium &  \\ 
K01829.01 & 15.6 & 2013/10/23  & low &  \\ 
K01830.01 & 14.2 & 2013/07/27  & medium & yes \\ 
K01833.01 & 13.8 & 2013/08/17  & medium &  \\ 
K01837.01 & 13.5 & 2013/08/18  & medium &  \\ 
K01838.01 & 14.4 & 2013/08/21  & medium &  \\ 
K01841.01 & 12.9 & 2013/08/17  & medium &  \\ 
K01842.01 & 14.1 & 2013/07/24  & medium &  \\ 
K01848.01 & 13.3 & 2013/08/13  & medium &  \\ 
K01849.01 & 14.3 & 2013/08/19  & low &  \\ 
K01851.01 & 14.6 & 2013/08/21  & medium &  \\ 
K01853.01 & 13.3 & 2013/08/14  & medium & yes \\ 
K01861.01 & 13.8 & 2013/07/28  & high & yes \\ 
K01864.01 & 13.2 & 2013/07/24  & high &  \\ 
K01870.01 & 14.1 & 2013/07/24  & medium &  \\ 
K01875.01 & 14.3 & 2013/07/28  & medium &  \\ 
K01876.01 & 14.7 & 2013/10/22  & low &  \\ 
K01877.01 & 13.1 & 2013/08/14  & medium &  \\ 
K01881.01 & 14.6 & 2013/08/18  & low &  \\ 
K01885.01 & 14.2 & 2013/08/13  & medium &  \\ 
K01898.01 & 13.1 & 2012/08/29 & medium ($i'$-band) &  \\ 
K01899.01 & 14.4 & 2013/07/27  & high & yes \\ 
K01902.01 & 14.0 & 2013/08/22  & medium &  \\ 
K01904.01 & 13.1 & 2013/07/27  & high &  \\ 
K01906.01 & 14.6 & 2013/10/22  & low &  \\ 
K01911.01 & 14.9 & 2013/08/22  & low &  \\ 
K01914.01 & 14.1 & 2013/07/29  & medium &  \\ 
K01920.01 & 14.4 & 2013/08/13  & low &  \\ 
K01928.01 & 13.4 & 2012/08/29 & low ($i'$-band) &  \\ 
K01934.01 & 14.2 & 2013/07/27  & medium &  \\ 
K01935.01 & 15.6 & 2013/08/17  & low &  \\ 
K01937.01 & 13.1 & 2013/08/14  & high &  \\ 
K01942.01 & 15.1 & 2013/08/21  & low &  \\ 
K01946.01 & 14.3 & 2013/08/15  & low &  \\ 
K01950.01 & 15.7 & 2013/08/22  & low & yes \\ 
K01958.01 & 15.0 & 2013/07/21  & medium &  \\ 
K01965.01 & 14.1 & 2013/08/17  & medium &  \\ 
K01967.01 & 14.0 & 2013/07/29  & medium &  \\ 
K01969.01 & 14.8 & 2013/08/21  & low &  \\ 
K01972.01 & 13.6 & 2013/08/15  & medium & yes \\ 
K01976.01 & 14.8 & 2013/08/21  & low &  \\ 
K01978.01 & 15.0 & 2012/09/13  & low &  \\ 
K01980.01 & 13.6 & 2013/08/15  & high &  \\ 
K01985.01 & 13.4 & 2013/07/24  & medium & yes \\ 
K01986.01 & 14.5 & 2013/08/21  & low &  \\ 
K01989.01 & 13.1 & 2013/08/14  & low & yes \\ 
K01990.01 & 14.3 & 2013/08/16  & medium &  \\ 
K01992.01 & 14.3 & 2013/08/17  & medium &  \\ 
K02005.01 & 14.4 & 2013/08/22  & low &  \\ 
K02007.01 & 13.0 & 2013/08/15  & medium &  \\ 
K02014.01 & 15.4 & 2013/10/23  & low & yes \\ 
K02018.01 & 14.3 & 2013/07/24  & medium &  \\ 
K02019.01 & 15.4 & 2013/08/22  & low & yes \\ 
K02023.01 & 14.9 & 2013/08/21  & low &  \\ 
K02024.01 & 14.4 & 2013/07/27  & medium &  \\ 
K02031.01 & 14.3 & 2013/08/21  & medium &  \\ 
K02032.01 & 12.0 & 2013/07/28  & high & yes \\ 
K02034.01 & 15.3 & 2013/08/22  & low &  \\ 
K02036.01 & 15.2 & 2013/08/22  & low &  \\ 
K02039.01 & 14.1 & 2013/08/14  & medium &  \\ 
K02043.01 & 13.9 & 2013/07/28  & high &  \\ 
K02052.01 & 15.2 & 2013/08/18  & low &  \\ 
K02055.01 & 14.3 & 2013/07/25  & low & yes \\ 
K02056.01 & 14.3 & 2013/08/16  & medium & yes \\ 
K02064.01 & 14.8 & 2013/08/21  & low &  \\ 
K02066.01 & 14.7 & 2013/08/22  & low &  \\ 
K02067.01 & 12.3 & 2013/07/24  & high & yes \\ 
K02069.01 & 13.6 & 2013/08/14  & medium & yes \\ 
K02075.01 & 12.1 & 2013/08/17  & high &  \\ 
K02081.01 & 13.1 & 2013/07/27  & high &  \\ 
K02083.01 & 13.4 & 2013/07/28  & medium & yes \\ 
K02084.01 & 14.8 & 2013/08/22  & low &  \\ 
K02095.01 & 14.5 & 2013/08/21  & medium &  \\ 
K02096.01 & 14.9 & 2013/08/19  & low & yes \\ 
K02097.01 & 14.5 & 2013/07/28  & low &  \\ 
K02098.01 & 13.7 & 2013/07/27  & medium & yes \\ 
K02099.01 & 14.8 & 2013/08/22  & low &  \\ 
K02100.01 & 14.4 & 2013/07/28  & medium & yes \\ 
K02102.01 & 14.9 & 2013/08/22  & low &  \\ 
K02109.01 & 13.9 & 2013/08/17  & high &  \\ 
K02114.01 & 14.7 & 2013/08/22  & low &  \\ 
K02115.01 & 15.8 & 2013/07/21  & low & yes \\ 
K02120.01 & 13.4 & 2013/07/25  & high &  \\ 
K02122.01 & 14.1 & 2013/07/28  & medium &  \\ 
K02123.01 & 14.2 & 2013/07/24  & medium &  \\ 
K02124.01 & 13.8 & 2013/08/15  & medium &  \\ 
K02130.01 & 15.1 & 2013/08/18  & low &  \\ 
K02131.01 & 14.5 & 2013/07/29  & low &  \\ 
K02132.01 & 14.3 & 2013/07/28  & medium &  \\ 
K02144.01 & 14.7 & 2013/08/21  & low &  \\ 
K02145.01 & 14.8 & 2013/08/22  & low &  \\ 
K02147.01 & 14.3 & 2013/08/16  & medium &  \\ 
K02148.01 & 13.1 & 2013/07/27  & medium &  \\ 
K02153.01 & 13.5 & 2013/07/27  & low &  \\ 
K02156.01 & 15.3 & 2013/08/18  & low & yes \\ 
K02160.01 & 14.7 & 2013/10/23  & low &  \\ 
K02168.01 & 14.7 & 2013/10/22  & low &  \\ 
K02172.01 & 14.9 & 2013/08/22  & low &  \\ 
K02174.01 & 15.2 & 2012/08/07  & low & yes \\ 
K02181.01 & 14.1 & 2013/07/25  & medium &  \\ 
K02185.01 & 14.2 & 2013/08/14  & medium &  \\ 
K02186.01 & 15.2 & 2013/08/18  & low &  \\ 
K02189.01 & 14.9 & 2013/08/21  & low &  \\ 
K02208.01 & 12.4 & 2013/08/16  & high &  \\ 
K02209.01 & 14.1 & 2013/08/13  & medium &  \\ 
K02218.01 & 14.3 & 2013/08/18  & low &  \\ 
K02225.01 & 14.0 & 2013/08/14  & medium &  \\ 
K02226.01 & 15.1 & 2013/08/22  & low &  \\ 
K02232.01 & 15.1 & 2013/08/22  & low &  \\ 
K02241.01 & 15.1 & 2013/08/18  & low &  \\ 
K02243.01 & 14.6 & 2013/08/22  & low &  \\ 
K02247.01 & 15.0 & 2013/07/27  & medium & yes \\ 
K02253.01 & 14.4 & 2013/08/13  & medium &  \\ 
K02256.01 & 15.3 & 2013/08/21  & low &  \\ 
K02259.01 & 15.1 & 2013/08/21  & low &  \\ 
K02261.01 & 13.8 & 2013/07/28  & high &  \\ 
K02263.01 & 14.3 & 2013/08/14  & medium &  \\ 
K02264.01 & 14.2 & 2013/08/17  & medium &  \\ 
K02266.01 & 15.6 & 2013/08/22  & low &  \\ 
K02269.01 & 15.0 & 2013/08/22  & low &  \\ 
K02278.01 & 13.6 & 2013/07/28  & high &  \\ 
K02280.01 & 15.7 & 2013/08/18  & low &  \\ 
K02282.01 & 14.0 & 2013/07/24  & medium &  \\ 
K02286.01 & 14.8 & 2013/08/21  & low &  \\ 
K02288.01 & 15.2 & 2013/08/22  & low &  \\ 
K02290.01 & 14.2 & 2013/08/15  & medium &  \\ 
K02293.01 & 14.6 & 2013/08/22  & medium &  \\ 
K02295.01 & 11.4 & 2013/07/29  & high & yes \\ 
K02296.01 & 14.4 & 2013/08/18  & medium &  \\ 
K02297.01 & 14.1 & 2013/07/25  & low &  \\ 
K02298.01 & 13.5 & 2013/08/16  & high & yes \\ 
K02299.01 & 15.1 & 2013/10/23  & low &  \\ 
K02304.01 & 15.1 & 2013/10/23  & low &  \\ 
K02305.01 & 15.5 & 2013/10/23  & low &  \\ 
K02306.01 & 14.2 & 2013/08/22  & low &  \\ 
K02310.01 & 14.4 & 2013/08/15  & medium &  \\ 
K02311.01 & 12.4 & 2013/07/29  & high &  \\ 
K02314.01 & 14.4 & 2013/07/27  & medium & yes \\ 
K02315.01 & 15.3 & 2013/08/22  & low &  \\ 
K02317.01 & 14.1 & 2013/07/27  & high & yes \\ 
K02318.01 & 14.1 & 2013/08/17  & medium &  \\ 
K02321.01 & 14.5 & 2013/08/22  & medium &  \\ 
K02325.01 & 14.1 & 2013/08/13  & medium &  \\ 
K02333.01 & 13.5 & 2012/08/31  & low ($i'$-band) &  \\ 
K02337.01 & 15.6 & 2013/08/22  & low &  \\ 
K02343.01 & 14.2 & 2013/08/14  & medium &  \\ 
K02344.01 & 15.3 & 2013/10/23  & low &  \\ 
K02351.01 & 14.7 & 2013/08/18  & low &  \\ 
K02354.01 & 14.3 & 2013/07/27  & high &  \\ 
K02356.01 & 14.3 & 2013/08/15  & medium &  \\ 
K02363.01 & 14.1 & 2013/07/24  & medium & yes$\dagger$ \\ 
K02371.01 & 14.8 & 2013/10/22  & low &  \\ 
K02373.01 & 14.5 & 2013/08/16  & medium &  \\ 
K02377.01 & 14.5 & 2013/07/24  & medium & yes \\ 
K02378.01 & 14.2 & 2013/07/29  & low &  \\ 
K02380.01 & 14.0 & 2013/07/24  & high & yes \\ 
K02386.01 & 14.7 & 2013/08/22  & low &  \\ 
K02392.01 & 14.8 & 2013/08/22  & low &  \\ 
K02393.01 & 14.6 & 2013/10/23  & low &  \\ 
K02400.01 & 15.4 & 2013/08/21  & low &  \\ 
K02402.01 & 14.2 & 2013/08/13  & low &  \\ 
K02403.01 & 12.7 & 2013/07/24  & high &  \\ 
K02411.01 & 15.2 & 2013/08/21  & low &  \\ 
K02415.01 & 15.5 & 2013/08/20  & low &  \\ 
K02417.01 & 15.9 & 2012/07/17  & low &  \\ 
K02418.01 & 15.0 & 2013/08/22  & low &  \\ 
K02419.01 & 15.1 & 2013/08/21  & low &  \\ 
K02421.01 & 14.0 & 2013/07/25  & low & yes \\ 
K02422.01 & 14.5 & 2013/10/23  & medium &  \\ 
K02423.01 & 14.5 & 2013/08/22  & low &  \\ 
K02430.01 & 14.1 & 2013/07/28  & high &  \\ 
K02437.01 & 14.5 & 2013/08/21  & low &  \\ 
K02444.01 & 15.5 & 2013/08/21  & low &  \\ 
K02448.01 & 15.1 & 2013/08/22  & low &  \\ 
K02462.01 & 11.7 & 2013/07/24  & high &  \\ 
K02467.01 & 14.4 & 2013/08/16  & medium &  \\ 
K02468.01 & 15.2 & 2013/10/22  & low &  \\ 
K02469.01 & 14.7 & 2013/08/18  & low & yes \\ 
K02471.01 & 14.3 & 2013/07/29  & low &  \\ 
K02474.01 & 13.9 & 2013/08/13  & medium & yes \\ 
K02483.01 & 14.3 & 2013/07/25  & medium &  \\ 
K02489.01 & 15.5 & 2013/08/18  & low &  \\ 
K02490.01 & 15.4 & 2013/08/22  & low &  \\ 
K02492.01 & 13.7 & 2013/08/17  & high &  \\ 
K02493.01 & 15.0 & 2013/08/22  & low & yes \\ 
K02504.01 & 15.7 & 2013/10/23  & low &  \\ 
K02507.01 & 15.7 & 2013/10/23  & low &  \\ 
K02508.01 & 14.9 & 2013/08/21  & low &  \\ 
K02516.01 & 13.1 & 2013/07/27  & high & yes \\ 
K02517.01 & 14.3 & 2013/07/29  & low &  \\ 
K02519.01 & 14.0 & 2013/08/13  & medium &  \\ 
K02523.01 & 15.6 & 2013/08/21  & low &  \\ 
K02529.01 & 15.4 & 2013/08/18  & low &  \\ 
K02531.01 & 15.1 & 2013/08/22  & low &  \\ 
K02536.01 & 14.2 & 2013/07/24  & low &  \\ 
K02542.01 & 14.8 & 2013/08/21  & low & yes \\ 
K02550.01 & 15.2 & 2013/08/22  & low &  \\ 
K02551.01 & 15.5 & 2013/10/23  & low & yes \\ 
K02552.01 & 14.3 & 2013/08/13  & medium &  \\ 
K02560.01 & 15.1 & 2013/08/22  & low &  \\ 
K02569.01 & 14.0 & 2013/07/27  & high &  \\ 
K02571.01 & 14.2 & 2013/08/13  & medium &  \\ 
K02573.01 & 15.6 & 2013/08/21  & low &  \\ 
K02577.01 & 13.4 & 2013/08/17  & medium &  \\ 
K02587.01 & 14.9 & 2013/10/23  & low &  \\ 
K02590.01 & 11.5 & 2013/08/16  & high &  \\ 
K02598.01 & 14.0 & 2013/08/13  & medium & yes \\ 
K02601.01 & 13.8 & 2013/08/13  & high & yes \\ 
K02607.01 & 14.3 & 2013/07/29  & medium &  \\ 
K02620.01 & 15.6 & 2013/10/22  & low &  \\ 
K02623.01 & 13.3 & 2013/08/17  & medium &  \\ 
K02626.01 & 15.2 & 2013/10/25  & low &  \\ 
K02627.01 & 14.4 & 2013/07/29  & low &  \\ 
K02634.01 & 15.7 & 2013/08/21  & low &  \\ 
K02635.01 & 12.7 & 2013/07/24  & high &  \\ 
K02636.01 & 12.1 & 2013/08/14  & high &  \\ 
K02649.01 & 14.4 & 2013/07/28  & medium &  \\ 
K02652.01 & 14.7 & 2013/08/18  & low &  \\ 
K02654.01 & 15.6 & 2013/10/22  & low &  \\ 
K02656.01 & 14.9 & 2013/10/23  & low &  \\ 
K02658.01 & 14.2 & 2013/08/14  & medium &  \\ 
K02659.01 & 12.7 & 2013/08/13  & medium &  \\ 
K02664.01 & 15.3 & 2013/08/18  & low & yes \\ 
K02666.01 & 15.1 & 2013/08/22  & low &  \\ 
K02672.01 & 11.7 & 2013/08/14  & high &  \\ 
K02674.01 & 13.2 & 2013/07/27  & high &  \\ 
K02675.01 & 12.2 & 2013/07/28  & high &  \\ 
K02677.01 & 13.3 & 2013/08/13  & high &  \\ 
K02678.01 & 11.6 & 2013/08/13  & high &  \\ 
K02679.01 & 13.3 & 2013/07/25  & high & yes \\ 
K02680.01 & 14.4 & 2013/07/24  & medium &  \\ 
K02681.01 & 15.7 & 2013/10/23  & low & yes \\ 
K02686.01 & 13.5 & 2013/07/25  & medium &  \\ 
K02687.01 & 10.0 & 2013/08/13  & high &  \\ 
K02690.01 & 13.2 & 2013/07/28  & high &  \\ 
K02693.01 & 12.8 & 2013/07/29  & high &  \\ 
K02696.01 & 12.9 & 2013/08/14  & high &  \\ 
K02698.01 & 13.8 & 2013/07/24  & high &  \\ 
K02700.01 & 14.9 & 2013/08/22  & low &  \\ 
K02704.01 & 17.0 & 2013/08/21  & low &  \\ 
K02705.01 & 14.3 & 2013/07/25  & high & yes \\ 
K02706.01 & 10.2 & 2013/08/14  & high &  \\ 
K02707.01 & 14.2 & 2013/07/24  & medium & yes \\ 
K02711.01 & 13.5 & 2013/07/29  & medium & yes \\ 
K02712.01 & 11.0 & 2013/07/29  & high &  \\ 
K02714.01 & 13.2 & 2013/08/18  & high &  \\ 
K02717.01 & 12.3 & 2013/08/14  & high &  \\ 
K02720.01 & 10.2 & 2013/07/29  & high &  \\ 
K02722.01 & 13.2 & 2013/08/14  & high & yes \\ 
K02723.01 & 14.3 & 2013/07/27  & medium &  \\ 
K02729.01 & 13.7 & 2013/08/18  & medium & yes \\ 
K02730.01 & 13.7 & 2013/07/28  & high &  \\ 
K02732.01 & 12.7 & 2013/08/15  & high &  \\ 
K02733.01 & 13.6 & 2013/08/13  & high &  \\ 
K02734.01 & 15.5 & 2013/08/21  & low &  \\ 
K02743.01 & 13.5 & 2013/08/14  & high & yes \\ 
K02748.01 & 13.8 & 2013/07/29  & low &  \\ 
K02750.01 & 15.7 & 2013/08/18  & low &  \\ 
K02753.01 & 13.4 & 2013/07/27  & high &  \\ 
K02754.01 &  & 2013/08/14  & high & yes \\ 
K02755.01 & 12.0 & 2013/07/28  & high &  \\ 
K02756.01 & 14.6 & 2013/10/23  & medium &  \\ 
K02757.01 & 14.4 & 2013/08/13  & low &  \\ 
K02759.01 & 14.0 & 2013/08/18  & medium &  \\ 
K02771.01 &  & 2013/08/16  & high & yes \\ 
K02775.01 & 14.7 & 2013/08/22  & low &  \\ 
K02778.01 & 15.1 & 2013/08/22  & low &  \\ 
K02779.01 & 14.8 & 2013/10/22  & low & yes \\ 
K02785.01 & 13.9 & 2013/08/16  & high &  \\ 
K02786.01 & 13.7 & 2013/08/15  & medium &  \\ 
K02789.01 & 14.3 & 2013/08/14  & medium &  \\ 
K02790.01 & 13.1 & 2013/07/24  & high &  \\ 
K02795.01 & 15.1 & 2013/08/17  & low &  \\ 
K02798.01 & 13.6 & 2013/07/29  & medium &  \\ 
K02801.01 & 10.6 & 2013/07/24  & high &  \\ 
K02803.01 & 12.1 & 2013/08/13  & high & yes \\ 
K02805.01 & 13.2 & 2013/08/17  & high &  \\ 
K02807.01 & 13.7 & 2013/07/28  & high & yes \\ 
K02812.01 & 14.2 & 2013/07/27  & high & yes \\ 
K02822.01 & 15.2 & 2013/08/18  & low &  \\ 
K02827.01 & 14.3 & 2013/08/13  & low &  \\ 
K02829.01 & 12.7 & 2013/08/14  & high &  \\ 
K02830.01 & 13.8 & 2013/07/24  & high &  \\ 
K02833.01 & 12.4 & 2013/07/27  & high &  \\ 
K02836.01 & 14.9 & 2013/10/22  & low & yes \\ 
K02837.01 & 13.1 & 2013/08/15  & high & yes \\ 
K02838.01 & 13.2 & 2013/08/13  & high & yes \\ 
K02840.01 & 13.7 & 2013/07/25  & high &  \\ 
K02842.01 & 15.5 & 2013/07/27  & high &  \\ 
K02848.01 & 12.3 & 2013/08/14  & high & yes \\ 
K02849.01 & 14.3 & 2013/08/16  & medium & \\ 
K02855.01 & 14.2 & 2013/08/13  & medium &  \\ 
K02857.01 & 13.5 & 2013/07/27  & medium &  \\ 
K02859.01 & 13.6 & 2013/08/16  & high & yes \\ 
K02866.01 & 14.0 & 2013/07/27  & medium &  \\ 
K02867.01 & 12.6 & 2013/08/14  & high &  \\ 
K02869.01 & 13.6 & 2013/07/27  & high & \\ 
K02882.01 & 15.0 & 2013/08/21  & low &  \\ 
K02883.01 & 15.4 & 2013/08/22  & low &  \\ 
K02890.01 & 15.0 & 2013/08/22  & low &  \\ 
K02904.01 & 12.5 & 2013/07/24  & low & yes \\ 
K02906.01 & 13.6 & 2013/07/25  & high &  \\ 
K02910.01 & 15.0 & 2013/08/19  & low & yes \\ 
K02913.01 & 12.7 & 2013/07/29  & high &  \\ 
K02914.01 & 12.1 & 2013/07/24  & high & yes \\ 
K02915.01 &  & 2013/07/29  & medium &  \\ 
K02923.01 & 14.4 & 2013/08/13  & low &  \\ 
K02924.01 & 14.3 & 2013/08/13  & medium &  \\ 
K02935.01 & 13.5 & 2013/07/24  & high &  \\ 
K02936.01 & 14.2 & 2013/07/27  & high &  \\ 
K02942.01 & 15.2 & 2013/08/18  & low &  \\ 
K02943.01 & 13.7 & 2013/08/13  & high &  \\ 
K02948.01 & 11.7 & 2013/08/14  & high &  \\ 
K02949.01 & 13.1 & 2013/07/29  & medium & yes \\ 
K02951.01 & 13.2 & 2013/07/25  & low &  \\ 
K02956.01 & 11.7 & 2013/07/28  & high &  \\ 
K02959.01 & 15.3 & 2013/08/22  & low &  \\ 
K02961.01 & 12.4 & 2013/08/16  & high &  \\ 
K02962.01 & 14.0 & 2013/07/25  & low & yes \\ 
K02963.01 & 14.4 & 2013/08/13  & medium &  \\ 
K02964.01 & 13.9 & 2013/07/24  & medium &  \\ 
K02967.01 & 15.3 & 2013/08/22  & low &  \\ 
K02968.01 &  & 2013/08/15  & high &  \\ 
K02970.01 & 12.7 & 2013/07/28  & high &  \\ 
K02971.01 & 12.6 & 2013/07/24  & high & yes \\ 
K02977.01 & 13.6 & 2013/08/15  & high &  \\ 
K02982.01 & 14.0 & 2013/08/15  & medium &  \\ 
K02984.01 & 12.9 & 2013/07/24  & high & yes \\ 
K02989.01 & 15.1 & 2013/08/18  & low &  \\ 
K02992.01 & 15.4 & 2013/10/23  & low &  \\ 
K02994.01 & 14.4 & 2013/08/16  & medium &  \\ 
K02998.01 & 15.6 & 2013/08/21  & low &  \\ 
K03004.01 & 14.7 & 2013/08/21  & medium &  \\ 
K03007.01 & 14.3 & 2013/08/14  & low &  \\ 
K03008.01 & 11.9 & 2013/07/27  & high &  \\ 
K03015.01 & 13.1 & 2013/08/16  & high &  \\ 
K03017.01 & 13.1 & 2013/08/15  & high &  \\ 
K03019.01 & 14.7 & 2013/10/22  & low &  \\ 
K03026.01 & 14.6 & 2013/08/22  & medium &  \\ 
K03029.01 & 14.7 & 2013/08/18  & low & yes \\ 
K03038.01 & 14.0 & 2013/07/25  & medium &  \\ 
K03039.01 & 13.7 & 2013/07/29  & high &  \\ 
K03041.01 & 14.0 & 2013/07/25  & medium & yes \\ 
K03048.01 & 14.4 & 2013/08/14  & medium &  \\ 
K03050.01 & 15.1 & 2013/08/18  & low &  \\ 
K03052.01 & 13.3 & 2013/07/29  & medium &  \\ 
K03053.01 & 15.5 & 2013/08/21  & low &  \\ 
K03056.01 & 15.6 & 2013/10/23  & low &  \\ 
K03057.01 & 15.6 & 2013/08/22  & low &  \\ 
K03060.01 & 12.7 & 2013/08/16  & high &  \\ 
K03061.01 & 14.0 & 2013/07/25  & medium &  \\ 
K03065.01 & 14.4 & 2013/07/24  & medium &  \\ 
K03068.01 & 14.7 & 2013/08/18  & medium &  \\ 
K03069.01 & 14.7 & 2013/08/19  & low & yes \\ 
K03071.01 & 14.1 & 2013/08/13  & medium &  \\ 
K03073.01 & 14.2 & 2013/08/14  & medium & yes \\ 
K03075.01 & 12.9 & 2013/07/28  & high &  \\ 
K03083.01 & 12.7 & 2013/08/13  & high &  \\ 
K03086.01 & 15.4 & 2013/08/22  & low &  \\ 
K03093.01 & 13.4 & 2013/07/29  & medium &  \\ 
K03097.01 & 11.8 & 2013/07/24  & high &  \\ 
K03105.01 & 15.0 & 2013/10/23  & low &  \\ 
K03109.01 & 13.5 & 2013/07/29  & medium &  \\ 
K03117.01 & 12.9 & 2013/08/13  & high &  \\ 
K03119.01 & 16.2 & 2013/08/22  & low &  \\ 
K03122.01 & 12.0 & 2013/08/15  & high &  \\ 
K03123.01 & 14.8 & 2013/08/19  & low &  \\ 
K03125.01 & 12.8 & 2013/08/15  & high &  \\ 
K03127.01 & 14.7 & 2013/08/22  & low &  \\ 
K03128.01 & 13.2 & 2013/08/13  & high &  \\ 
K03130.01 & 13.5 & 2013/07/24  & medium &  \\ 
K03140.01 & 15.0 & 2013/08/22  & low &  \\ 
K03142.01 & 15.1 & 2013/08/22  & low &  \\ 
K03145.01 & 15.3 & 2013/08/18  & low &  \\ 
K03146.01 & 13.9 & 2013/08/15  & medium &  \\ 
K03147.01 & 14.1 & 2013/07/29  & low &  \\ 
K03158.01 &  & 2013/07/21  & high & yes \\ 
K03179.01 & 10.5 & 2013/07/24  & high &  \\ 
K03184.01 &  & 2013/08/13  & high &  \\ 
K03190.01 & 11.1 & 2013/07/27  & high & yes \\ 
K03194.01 & 11.4 & 2013/08/17  & high &  \\ 
K03196.01 & 11.4 & 2013/07/28  & high &  \\ 
K03203.01 & 11.7 & 2013/07/24  & high &  \\ 
K03204.01 &  & 2013/08/16  & high &  \\ 
K03208.01 & 11.8 & 2013/07/29  & high &  \\ 
K03209.01 & 11.8 & 2013/08/13  & high &  \\ 
K03220.01 & 12.0 & 2013/08/17  & high &  \\ 
K03232.01 & 12.1 & 2013/07/27  & high &  \\ 
K03234.01 & 12.2 & 2013/08/14  & high &  \\ 
K03236.01 & 12.2 & 2013/07/27  & high &  \\ 
K03237.01 & 12.1 & 2013/07/28  & high &  \\ 
K03242.01 & 12.4 & 2013/07/28  & high &  \\ 
K03245.01 & 12.3 & 2013/07/25  & high & yes \\ 
K03246.01 & 12.1 & 2013/08/16  & high &  \\ 
K03255.01 & 14.0 & 2013/07/27  & high & yes \\ 
K03259.01 & 15.4 & 2013/08/22  & low &  \\ 
K03260.01 & 14.5 & 2013/08/18  & low &  \\ 
K03261.01 & 14.2 & 2013/08/17  & medium &  \\ 
K03277.01 & 12.9 & 2013/07/24  & high & yes \\ 
K03278.01 & 14.5 & 2013/08/22  & medium &  \\ 
K03284.01 & 13.8 & 2013/08/22  & low & yes \\ 
K03288.01 & 14.0 & 2013/07/24  & high & yes \\ 
K03296.01 & 14.2 & 2013/07/27  & high &  \\ 
K03301.01 & 14.0 & 2013/08/14  & medium &  \\ 
K03308.01 & 13.9 & 2013/07/29  & medium &  \\ 
K03309.01 & 14.4 & 2013/08/17  & medium & yes \\ 
K03310.01 & 13.3 & 2013/08/18  & high &  \\ 
K03311.01 & 14.0 & 2013/08/15  & medium &  \\ 
K03315.01 & 13.8 & 2013/08/15  & high &  \\ 
K03324.01 & 15.7 & 2013/10/23  & low & yes \\ 
K03334.01 & 15.0 & 2013/08/13  & high &  \\ 
K03339.01 & 14.4 & 2013/10/22  & low & yes \\ 
K03340.01 & 13.6 & 2013/07/28  & high &  \\ 
K03343.01 & 14.8 & 2013/08/22  & low &  \\ 
K03345.01 & 14.5 & 2013/08/15  & medium &  \\ 
K03346.01 & 13.2 & 2013/07/28  & high &  \\ 
K03353.01 & 14.3 & 2013/07/24  & low &  \\ 
K03363.01 & 13.6 & 2013/08/17  & high &  \\ 
K03365.01 & 14.2 & 2013/08/17  & medium &  \\ 
K03371.01 & 13.5 & 2013/08/17  & high &  \\ 
K03374.01 & 14.3 & 2013/07/27  & high &  \\ 
K03377.01 & 14.9 & 2013/08/22  & low & yes \\ 
K03379.01 & 15.1 & 2013/08/22  & low &  \\ 
K03384.01 & 13.0 & 2013/07/27  & high &  \\ 
K03385.01 & 15.0 & 2013/08/21  & low &  \\ 
K03389.01 & 14.4 & 2013/07/29  & low &  \\ 
K03398.01 & 13.4 & 2013/07/24  & high &  \\ 
K03401.01 & 14.2 & 2013/07/28  & medium & yes \\ 
K03403.01 & 12.9 & 2013/08/17  & high &  \\ 
K03408.01 & 14.6 & 2013/10/23  & medium &  \\ 
K03414.01 & 14.9 & 2013/08/22  & low &  \\ 
K03419.01 & 12.9 & 2013/07/29  & medium &  \\ 
K03423.01 & 14.1 & 2013/07/28  & medium &  \\ 
K03425.01 & 13.0 & 2013/07/24  & high &  \\ 
K03429.01 & 15.1 & 2013/10/23  & low &  \\ 
K03430.01 & 15.4 & 2013/10/23  & low &  \\ 
K03437.01 & 14.4 & 2013/08/13  & medium &  \\ 
K03438.01 & 13.6 & 2013/08/14  & high &  \\ 
K03439.01 & 14.0 & 2013/07/24  & medium & yes \\ 
K03440.01 & 15.4 & 2013/08/18  & low &  \\ 
K03444.01 & 13.0 & 2013/07/25  & high & yes \\ 
K03448.01 & 14.7 & 2013/08/22  & low &  \\ 
K03451.01 & 13.9 & 2013/08/17  & medium &  \\ 
K03456.01 & 12.8 & 2013/08/13  & high &  \\ 
K03459.01 & 15.0 & 2013/08/21  & low & yes \\ 
K03460.01 & 14.3 & 2013/07/27  & high & yes \\ 
K03468.01 & 14.1 & 2013/07/24  & medium & yes \\ 
K03474.01 & 14.0 & 2013/07/24  & medium &  \\ 
K03481.01 & 14.3 & 2013/08/15  & medium &  \\ 
K03484.01 & 15.0 & 2013/08/21  & low &  \\ 
K03486.01 & 14.1 & 2013/07/24  & medium & yes \\ 
K03496.01 & 13.8 & 2013/08/14  & high &  \\ 
K03497.01 & 13.0 & 2013/10/24  & medium & yes \\ 
K03500.01 & 13.0 & 2013/07/25  & high & yes \\ 
K03503.01 & 13.7 & 2013/07/29  & medium &  \\ 
K03506.01 & 12.8 & 2013/07/24  & high &  \\ 
K03528.01 & 12.6 & 2013/07/29  & high &  \\ 
K03541.01 & 13.6 & 2013/07/28  & high &  \\ 
K03557.01 & 13.2 & 2013/07/29  & high &  \\ 
K03573.01 & 13.5 & 2013/08/16  & high &  \\ 
K03583.01 & 12.5 & 2013/07/24  & high &  \\ 
K03602.01 & 13.7 & 2013/08/14  & medium &  \\ 
K03608.01 & 13.4 & 2013/08/14  & high &  \\ 
K03627.01 & 15.0 & 2013/08/22  & low &  \\ 
K03663.01 & 12.4 & 2013/08/18  & high &  \\ 
K03680.01 & 14.3 & 2013/08/14  & low &  \\ 
K03681.01 &  & 2013/07/24  & high &  \\ 
K03683.01 & 11.9 & 2013/08/17  & high &  \\ 
K03689.01 & 13.9 & 2013/08/13  & low &  \\ 
K03721.01 & 13.9 & 2013/07/29  & medium &  \\ 
K03728.01 & 12.2 & 2013/08/15  & high &  \\ 
K03762.01 & 14.6 & 2013/10/23  & low &  \\ 
K03780.01 & 13.9 & 2013/08/14  & medium &  \\ 
K03782.01 & 13.2 & 2013/08/16  & high &  \\ 
K03787.01 & 13.7 & 2013/08/14  & high &  \\ 
K03811.01 & 13.7 & 2013/07/28  & high &  \\ 
K03835.01 & 12.5 & 2013/07/24  & high &  \\ 
K03837.01 & 14.0 & 2013/08/15  & medium &  \\ 
K03848.01 & 14.0 & 2013/08/18  & medium &  \\ 
K03864.01 & 12.6 & 2013/07/28  & high &  \\ 
K03875.01 & 13.3 & 2013/08/17  & high &  \\ 
K03876.01 & 12.4 & 2013/07/29  & high &  \\ 
K03891.01 & 13.4 & 2013/07/27  & high & yes \\ 
K03892.01 & 12.5 & 2013/08/17  & high &  \\ 
K03893.01 & 15.2 & 2013/08/21  & low &  \\ 
K03907.01 & 12.5 & 2013/07/27  & high & yes \\ 
K03908.01 & 11.9 & 2013/07/24  & high &  \\ 
K03911.01 & 14.3 & 2013/08/15  & medium &  \\ 
K03916.01 & 15.0 & 2013/08/18  & low &  \\ 
K03925.01 & 13.8 & 2013/08/18  & medium &  \\ 
K03935.01 & 14.6 & 2013/08/22  & medium &  \\ 
K03943.01 & 12.5 & 2013/08/15  & high &  \\ 
K03946.01 & 13.1 & 2013/08/13  & high & yes \\ 
K03966.01 & 14.0 & 2013/07/29  & low &  \\ 
K03984.01 & 14.2 & 2013/07/29  & medium &  \\ 
K03991.01 & 12.9 & 2013/08/13  & high &  \\ 
K04004.01 & 12.5 & 2013/07/27  & high & yes \\ 
K04005.01 & 14.3 & 2013/07/24  & low &  \\ 
K04014.01 & 13.5 & 2013/08/16  & high &  \\ 
K04016.01 & 13.6 & 2013/07/27  & medium &  \\ 
K04021.01 & 12.5 & 2013/08/16  & high & yes \\ 
K04032.01 & 12.4 & 2013/07/28  & high &  \\ 
K04037.01 & 12.7 & 2013/07/24  & high &  \\ 
K04051.01 & 14.8 & 2013/08/21  & low &  \\ 
K04053.01 & 12.6 & 2013/07/25  & high & yes \\ 
K04054.01 & 14.3 & 2013/07/28  & medium &  \\ 
K04056.01 & 14.1 & 2013/07/25  & medium &  \\ 
K04060.01 & 13.7 & 2013/08/14  & high &  \\ 
K04066.01 & 13.6 & 2013/08/15  & high &  \\ 
K04067.01 & 14.0 & 2013/07/29  & medium &  \\ 
K04068.01 & 15.1 & 2013/08/22  & low &  \\ 
K04069.01 & 14.5 & 2013/08/17  & medium &  \\ 
K04072.01 & 13.3 & 2013/07/27  & high &  \\ 
K04097.01 & 13.0 & 2013/07/29  & high &  \\ 
K04098.01 & 13.5 & 2013/08/15  & low & yes \\ 
K04109.01 & 14.2 & 2013/07/27  & high &  \\ 
K04123.01 & 13.3 & 2013/07/25  & medium &  \\ 
K04129.01 & 13.2 & 2013/08/15  & high &  \\ 
K04136.01 & 13.8 & 2013/07/24  & medium &  \\ 
K04145.01 & 14.1 & 2013/07/27  & medium & yes \\ 
K04146.01 & 13.3 & 2013/08/15  & high &  \\ 
K04149.01 & 14.1 & 2013/10/22  & low & yes \\ 
K04150.01 & 14.5 & 2013/10/23  & medium &  \\ 
K04152.01 & 12.9 & 2013/08/14  & high &  \\ 
K04153.01 & 14.5 & 2013/10/23  & low &  \\ 
K04156.01 & 13.6 & 2013/07/28  & high &  \\ 
K04157.01 & 12.6 & 2013/08/13  & high &  \\ 
K04159.01 & 14.2 & 2013/08/13  & medium &  \\ 
K04160.01 & 13.2 & 2013/08/13  & high &  \\ 
K04166.01 & 15.0 & 2013/08/18  & low & yes \\ 
K04169.01 & 14.4 & 2013/08/22  & medium &  \\ 
K04184.01 & 13.2 & 2013/07/24  & high &  \\ 
K04188.01 & 13.0 & 2013/07/27  & high &  \\ 
K04190.01 & 13.4 & 2013/08/18  & medium &  \\ 
K04194.01 & 15.0 & 2013/08/21  & low & yes \\ 
K04198.01 & 12.5 & 2013/07/29  & high &  \\ 
K04199.01 & 14.0 & 2013/07/27  & medium &  \\ 
K04205.01 & 14.2 & 2013/07/28  & medium & yes \\ 
K04208.01 & 15.3 & 2013/08/17  & low & yes \\ 
K04209.01 & 15.7 & 2013/08/19  & low & yes \\ 
K04212.01 & 13.8 & 2013/07/25  & high &  \\ 
K04215.01 & 13.7 & 2013/07/24  & medium &  \\ 
K04226.01 & 12.6 & 2013/07/24  & high & yes \\ 
K04230.01 & 13.7 & 2013/07/24  & high &  \\ 
K04235.01 & 13.1 & 2013/08/17  & high &  \\ 
K04242.01 & 14.2 & 2013/08/15  & low &  \\ 
K04245.01 & 13.4 & 2013/08/18  & high &  \\ 
K04251.01 & 13.6 & 2013/08/14  & high &  \\ 
K04252.01 & 13.4 & 2013/08/17  & high &  \\ 
K04253.01 & 14.1 & 2013/08/13  & low &  \\ 
K04269.01 & 12.9 & 2013/08/13  & high &  \\ 
K04273.01 & 12.5 & 2013/08/14  & high &  \\ 
K04274.01 & 15.1 & 2013/08/22  & low & yes \\ 
K04276.01 & 12.4 & 2013/07/29  & high &  \\ 
K04287.01 & 11.1 & 2013/08/17  & high & yes \\ 
K04288.01 & 12.3 & 2013/07/29  & high &  \\ 
K04292.01 & 12.7 & 2013/08/15  & high & yes$\dagger$ \\ 
K04295.01 & 14.8 & 2013/08/18  & medium &  \\ 
K04296.01 & 12.9 & 2013/07/25  & high &  \\ 
K04300.01 & 15.7 & 2013/10/23  & low &  \\ 
K04301.01 & 12.9 & 2013/07/27  & high &  \\ 
K04302.01 & 13.5 & 2013/07/28  & high &  \\ 
K04304.01 & 13.9 & 2013/08/16  & medium &  \\ 
K04312.01 & 13.7 & 2013/08/17  & high &  \\ 
K04313.01 & 14.1 & 2013/08/13  & medium & yes \\ 
K04318.01 & 12.5 & 2013/08/14  & high &  \\ 
K04320.01 & 14.0 & 2013/07/29  & medium &  \\ 
K04325.01 & 15.5 & 2013/10/23  & low &  \\ 
K04327.01 & 14.8 & 2013/10/23  & low &  \\ 
K04329.01 & 11.9 & 2013/08/18  & high & yes \\ 
K04331.01 & 13.0 & 2013/07/29  & medium & yes \\ 
K04335.01 & 13.9 & 2013/08/16  & high &  \\ 
K04337.01 & 14.5 & 2013/10/23  & medium &  \\ 
K04341.01 & 13.1 & 2013/07/27  & high &  \\ 
K04347.01 & 14.6 & 2013/08/22  & low &  \\ 
K04356.01 & 15.5 & 2013/08/21  & low &  \\ 
K04367.01 & 13.8 & 2013/07/24  & medium &  \\ 
K04374.01 & 13.9 & 2013/07/28  & high &  \\ 
K04382.01 & 14.0 & 2013/08/14  & high &  \\ 
K04383.01 & 13.7 & 2013/08/16  & high &  \\ 
K04389.01 & 14.8 & 2013/08/21  & low & yes \\ 
K04390.01 & 16.1 & 2013/08/18  & low &  \\ 
K04393.01 & 15.0 & 2013/08/21  & low &  \\ 
K04395.01 & 15.4 & 2013/08/21  & low &  \\ 
K04399.01 & 11.8 & 2013/07/24  & high & yes \\ 
K04400.01 & 13.6 & 2013/07/24  & medium &  \\ 
K04403.01 & 15.5 & 2013/08/21  & low &  \\ 
K04407.01 & 11.0 & 2013/07/24  & high & yes \\ 
K04409.01 & 12.3 & 2013/07/24  & high & yes \\ 
K04411.01 & 13.2 & 2013/07/28  & high &  \\ 
K04423.01 & 13.0 & 2013/07/24  & high &  \\ 
K04427.01 & 15.0 & 2013/08/22  & low &  \\ 
K04428.01 & 14.4 & 2013/07/24  & medium &  \\ 
K04431.01 & 12.5 & 2013/08/14  & high &  \\ 
K04435.01 & 13.4 & 2013/07/24  & high &  \\ 
K04443.01 & 13.7 & 2013/08/13  & high & yes \\ 
K04446.01 & 12.2 & 2013/08/15  & high &  \\ 
K04456.01 & 14.9 & 2013/08/18  & low &  \\ 
K04457.01 & 13.7 & 2013/08/17  & high &  \\ 
K04463.01 & 14.6 & 2013/07/27  & medium & yes \\ 
K04474.01 & 15.6 & 2013/08/21  & low &  \\ 
K04478.01 & 14.3 & 2013/07/29  & medium &  \\ 
K04482.01 & 12.5 & 2013/07/24  & high &  \\ 
K04495.01 & 15.2 & 2013/08/22  & low & yes \\ 
K04505.01 & 13.5 & 2013/07/25  & high &  \\ 
K04510.01 & 14.2 & 2013/08/15  & medium &  \\ 
K04520.01 & 14.0 & 2013/08/13  & medium &  \\ 
K04523.01 & 14.6 & 2013/08/21  & medium & yes \\ 
K04530.01 & 14.0 & 2013/08/18  & medium &  \\ 
K04541.01 & 15.2 & 2013/08/22  & low &  \\ 
K04554.01 & 14.4 & 2013/07/28  & medium &  \\ 
K04556.01 & 11.8 & 2013/08/17  & high &  \\ 
K04558.01 & 14.4 & 2013/07/29  & low &  \\ 
K04561.01 & 14.0 & 2013/08/14  & high &  \\ 
K04563.01 & 14.3 & 2013/08/14  & medium &  \\ 
K04567.01 & 13.6 & 2013/07/24  & high & yes \\ 
K04570.01 & 14.5 & 2013/07/28  & medium &  \\ 
K04575.01 & 13.0 & 2013/08/16  & high & yes \\ 
K04580.01 & 12.8 & 2013/08/13  & high & yes \\ 
K04581.01 & 13.4 & 2013/07/24  & high &  \\ 
K04582.01 & 11.6 & 2013/07/27  & high & yes \\ 
K04586.01 & 14.2 & 2013/08/17  & medium &  \\ 
K04588.01 & 12.8 & 2013/08/15  & medium &  \\ 
K04591.01 & 15.0 & 2013/08/18  & low &  \\ 
K04601.01 & 13.6 & 2013/07/28  & high &  \\ 
K04603.01 & 13.5 & 2013/08/14  & high &  \\ 
K04605.01 & 13.3 & 2013/08/18  & high &  \\ 
K04613.01 & 13.7 & 2013/08/13  & high &  \\ 
K04614.01 & 13.7 & 2013/07/24  & medium &  \\ 
K04617.01 & 13.5 & 2013/08/16  & high &  \\ 
K04627.01 & 14.8 & 2013/08/18  & low &  \\ 
K04633.01 & 13.6 & 2013/08/14  & medium &  \\ 
K04634.01 & 13.5 & 2013/07/24  & high & yes \\ 
K04637.01 & 10.3 & 2013/08/15  & high &  \\ 
K04644.01 & 14.8 & 2013/08/18  & medium &  \\ 
K04651.01 & 13.6 & 2013/07/25  & high & yes \\ 
K04656.01 & 13.7 & 2013/07/29  & medium & yes \\ 
K04657.01 & 13.0 & 2013/07/29  & high & yes \\ 
K04663.01 & 12.5 & 2013/08/13  & high &  \\ 
K04674.01 & 13.4 & 2013/07/24  & high &  \\ 
K04680.01 & 15.3 & 2013/08/18  & low &  \\ 
K04691.01 & 14.3 & 2013/08/14  & medium &  \\ 
K04693.01 & 13.2 & 2013/08/13  & medium &  \\ 
K04699.01 & 12.8 & 2013/08/13  & high & yes \\ 
K04706.01 & 13.5 & 2013/07/28  & medium &  \\ 
K04709.01 & 15.7 & 2013/08/21  & low &  \\ 
K04737.01 & 14.5 & 2013/10/23  & medium &  \\ 
K04744.01 & 13.6 & 2013/07/28  & medium &  \\ 
K04747.01 & 14.8 & 2013/08/22  & low &  \\ 
K04754.01 & 15.1 & 2013/08/22  & low &  \\ 
K04763.01 & 13.9 & 2013/07/28  & high &  \\ 
K04768.01 & 15.4 & 2013/10/23  & low & yes \\ 
K04772.01 & 15.3 & 2013/08/18  & low &  \\ 
K04773.01 & 13.4 & 2013/07/27  & high &  \\ 
K04775.01 & 12.7 & 2013/08/14  & high &  \\ 
K04776.01 & 14.6 & 2013/08/21  & medium &  \\ 
K04782.01 & 15.6 & 2013/08/21  & low &  \\ 
K04792.01 & 14.0 & 2013/08/15  & medium & yes \\ 
K04797.01 & 15.3 & 2013/08/22  & low & yes \\ 
K04801.01 & 15.0 & 2013/10/23  & low &  \\ 
K04811.01 & 13.6 & 2013/08/15  & medium &  \\ 
K04812.01 & 15.5 & 2013/10/23  & low & yes \\ 
K04813.01 & 13.3 & 2013/07/28  & low & yes \\ 
K04818.01 & 12.9 & 2013/08/15  & high &  \\ 
K04822.01 & 13.3 & 2013/08/16  & high &  \\ 
K04823.01 & 12.5 & 2013/08/17  & high & yes \\ 
K04826.01 & 14.4 & 2013/07/29  & low &  \\ 
K04829.01 & 14.0 & 2013/08/18  & medium &  \\ 
K04833.01 & 13.4 & 2013/08/14  & high &  \\ 
K04834.01 & 12.8 & 2013/08/14  & high &  \\ 
K04839.01 & 13.3 & 2013/07/25  & medium &  \\ 
K04849.01 & 14.0 & 2013/07/25  & medium &  \\ 
K04850.01 & 13.9 & 2013/07/25  & medium &  \\ 
K04852.01 & 14.3 & 2013/08/15  & medium &  \\ 
K04854.01 & 14.1 & 2013/08/15  & medium &  \\ 
K04855.01 & 14.4 & 2013/08/16  & medium &  \\ 
K04871.01 & 12.9 & 2013/10/25  & medium & yes \\ 
K04875.01 & 15.2 & 2013/08/22  & low &  \\ 
K04877.01 & 14.3 & 2013/08/14  & medium &  \\ 
K04887.01 & 12.7 & 2013/07/28  & high &  \\ 
K04891.01 & 15.5 & 2013/08/22  & low &  \\ 
K04892.01 & 14.0 & 2013/08/15  & medium &  \\ 
K04907.01 & 13.6 & 2013/08/13  & high &  \\ 
K04914.01 & 14.0 & 2013/08/15  & medium &  \\
\enddata
\label{tab:obs_table}
\tablenotetext{$^*$}{All observations taken in the LP600 filter unless noted by ($i'$-band).}
\tablenotetext{$\dagger$}{Companion originally discovered in NIRC2 image, see Table \ref{tab:table_keck_nirc2}.}
\end{deluxetable*}








\end{document}